\documentclass[12pt]{article}
\usepackage{amsmath}
\usepackage{graphicx}
\date{}
\begin{document}
\title{A review of the neutrino emission processes in the late stages of the stellar evolutions}
\author{{\bf Indranath Bhattacharyya}\vspace{0.2cm}\\
Department of Mathematics\\
Barasat Government College\\10 K.N.C Road, Barasat (North 24 Parganas)\\ Kolkata-700124, West Bengal, India
\\E-mail :
$i_{-}bhattacharyya@hotmail.com$\\}\vskip .1in
\maketitle
\begin{abstract}
In this paper the neutrino emission processes being supposed to be the main sources of energy loss in the stellar
core in the later stages of stellar evolution are reviewed. All the calculations are carried out in the framework of electro-weak theory based on the Standard Model. It is considered that the neutrino has a little mass, which is very much consistent with the phenomenological evidences and presupposes a minimal extension of the Standard Model. All three neutrinos (i.e., electron neutrino, muon neutrino and tau neutrino) are taken into account in the calculations. It is evident that the strong magnetic field, present in the degenerate stellar objects such as neutron stars, has remarkable influence on some neutrino emission processes. The intensity of such magnetic field is very close to the critical value ($H_{c}=4.414\times 10^{13}$ G) and sometimes exceeds it. In this paper the region of dominance of different neutrino emission processes in absence of magnetic field is picturized. The region of importance of the neutrino emission processes in presence of a strong magnetic field is indicated by a picture. The study reveals the significant contributions of some neutrino emission processes, both in absence and in presence of a strong magnetic field in the later stages of stellar evolution.
\vspace{0.5cm}\\Keywords: Neutrino emission processes; late stages of stellar evolution; electro-weak interaction; scattering cross section; neutrino energy loss rate.
\end{abstract}
{\Large \bf Introduction:}\vspace{0.2cm}\\
The life span of a star is governed mostly by the mode of energy
loss in it. In the later phases of stellar evolution the life time
of stars depends mainly due to the emission of neutrinos via a
number of processes. The effect of the neutrino emission
processes, that are so important in the later stages, is
suppressed by the electromagnetic emission in the early stages of
stellar evolution. In the stellar objects, for examples, white
dwarves, neutron stars etc. neutrinos correspond to a sink of
energy coming out during their cooling period. The neutrino
emission occurs when the core of stars collapses in the high
temperature and density. In fact, the neutrino emission processes
draw attention of the scientists and researchers because of their
significant role to carry away the energy from the stellar core,
particularly in the later phases. In 1940 Gamow and Schoenberg
\cite{Gamow} speculated that the neutrino emission might be
significant during the collapse of evolved stars. They calculated
the energy losses through the neutrinos produced in the reactions
between free electrons and oxygen nuclei and suggested that such
energy losses could cause a complete collapse of the star within a
time period of half an hour. They named such interaction as URCA
process. At a very high temperature and density, which exist in
the interior of the contracting stars during the late stages, the
URCA process may cause the emission of large number of neutrinos.
These neutrinos penetrating almost without difficulty the body of
the star, must carry away quite a large amount of energy and
prevent the central temperature from rising above a certain limit,
which would result in a {\it catastrophic} collapse. Pontecorvo
\cite{Pontecorvo} predicted that the neutrino-antineutrino pair
might be formed as a result of bremsstrahlung process, i.e., when
the electron would collide with a heavy nucleus. He first
emphasized that the existence of a direct weak interaction between
low-energy neutrinos and electrons would have interesting and
profound implications for stellar evolution. Later, this
bremsstrahlung process was developed in the non-relativistic as
well as relativistic situations. Chiu and his collaborators
\cite{Chiu1960, Chiu1961, Chiu} considered a number of neutrino
emission processes to carry out the calculations in order to
obtain the neutrino luminosity of hot stars. Their works included
some neutrino emission processes and indicated the importance of
those processes during stellar evolution. After that many attempts
were made to study other neutrino emission processes in the
context of stellar evolution. In 1961 Matinyan and Tsilosani
\cite{Matinyan} investigated a new kind of mechanism in which
neutrino-antineutrino pair would be emitted by the interaction of
the photon with nucleus. In 1967 Landstreet \cite{Landstreet}
studied neutrino emission by the synchrotron process. That is very
much analogous to the electromagnetic synchrotron radiation.
Against this backdrop the need was felt for a detailed study of
all neutrino emission processes and the importance of their role
in stellar evolution.
\\\indent It is known that the stellar matter (even under some
extreme conditions as in white dwarves or in neutron stars) is
almost transparent to the neutrinos, in contrast to its behavior
with respect to photons. The neutrino emission is quite dissimilar
to the other mechanisms of energy transport from the core of the
star. The other mechanisms, such as electromagnetic radiations,
require the transport of internal energy to the surface to get
radiated and as a result the rate of energy loss is related to the
temperature gradient of the star. It has been observed that when
stellar core contracts considerably the neutrino emission
increases and the neutrinos, created in the stellar core, directly
come out of the star. This happens owing to its large mean free
path; since the cross-section of the interaction of neutrino with
charged particles is very low $(<10^{-44}$ $cm^{2})$.
Consequently, the weakness of the neutrino coupling with matter
leads to a situation where the very weakly emitted neutrino bears
more significance than emitted photons. In the higher density the
mean free path of the neutrinos much exceeds typical stellar
radii. As the neutrinos emerge directly from the point of origin,
the energy outflow does not depend upon the temperature gradient,
but is given directly by the rate at which these are produced. In
the astrophysical perspective it is very important to study those
neutrino emission processes as the energy loss mechanisms from
stellar objects in the later stages. There is a number of neutrino
emission processes which are supposed to play a very important
role in this regard. Earlier all those processes were studied
mostly in the framework of the current current coupling theory. In
1958 Feynman and Gell-Mann \cite{Feynman} as well as Sudarshan and
Marshak \cite{Sudarshan} proposed the universal V-A interaction
theory, according to which the weak interactions would be, in
fact, a mixture of vector and axial vector interactions. This
theory could explain the electron-neutrino elastic scattering
characterized by the strength of the Fermi coupling constant
$G_{F}$. The noticeable property of the neutrino is the weakness
of its interactions with other particles. These spectacular
neutrino interactions are characterized by the Fermi coupling
constant that measures the size of the matrix element for the
various processes. The development of this beautiful V-A theory
helped the scientists and researchers to explain those processes
in the low energy region. But V-A theory failed to explain the
high energy phenomena and the neutral current in weak interaction.
Introducing this neutral current a more generalized theory called
electro-weak theory was developed by Salam \cite{Salam}, Glashow
\cite{Glashow} and Weinberg \cite{Weinberg}. This theory succeeded
to unify electromagnetic and weak interactions and became an
important ingredient of the Standard Model \cite{Weinberg2004}.
According to this elegant theory the neutrino pair emission takes
place through the exchange of neutral $Z$-boson or charged
$W$-boson depending upon the charge conservation. In 1972 Dicus
\cite{Dicus1972} calculated a number of neutrino emission
processes in the framework of this electro-weak theory and
obtained the analytical expression for energy loss rate in
different regions to study their role in the astrophysical
scenario. His work clearly indicates the importance of those
neutrino emission processes to carry away the energy from the
stars. It may be mentioned that another theory which was advanced
by Bandyopadhyay and Raychaudhuri is photon neutrino weak coupling
theory \cite{Kuchowicz}. Some processes, such as, neutrino
synchrotron process \cite{Raychaudhuri3}, plasma neutrino process
\cite{Raychaudhuri1970} etc. give some important astrophysical
findings (specially in the cooling of white dwarves and neutron
stars etc.) in the framework of this photon neutrino weak coupling
theory as the cross-section is much smaller in the high energy
region than that obtained in the electro-weak theory.
Unfortunately, this photon neutrino weak coupling theory has not
been verified in the laboratory.\\\indent In the earlier works the
neutrino emissions were believed to be governed by three important
processes which are given as follows:
$$(i)\hspace{0.5cm}\Gamma\rightarrow\nu+\overline{\nu}\hspace{6.2cm}(plasma\hspace{0.2cm}
neutrino\hspace{0.2cm}process),$$
$$(ii)\hspace{0.5cm}\gamma+e^{-}\rightarrow e^{-}+\nu+\overline{\nu}\hspace{4.5cm}
(photo-neutrino \hspace{0.2cm}process),$$ and
$$(iii)\hspace{0.5cm}e^{+}+e^{-}\rightarrow\nu+\overline{\nu}\hspace{5cm}
(pair\hspace{0.2cm}annihilation\hspace{0.2cm}process).$$
 There exist a few more, such as, neutrino
bremsstrahlung process
($e^{-}+Z\longrightarrow^{magnetic\hspace{0.2cm}field}
e^{-}+Z+\nu+\overline{\nu}$), photon-photon scattering
($\gamma+\gamma\longrightarrow \nu+\overline{\nu}$
\cite{Dicus1972, Dicus1993, Abbasabadi1998, Abbasabadi2000};
$\gamma+\gamma\longrightarrow \gamma+\nu+\overline{\nu}$
\cite{Dicus1997, Dicus1998}) etc. that may also have some
significant effects under certain circumstances. In this paper we
consider the following four important neutrino emission processes.
$$(1)\hspace{0.5cm}\gamma+Z\longrightarrow
\gamma+Z+\nu+\overline{\nu}\hspace{2.8cm}(photo-coulomb\hspace{0.2cm}neutrino
\hspace{0.2cm}process),$$
$$(2)\hspace{0.5cm}e^{-}+e^{-}\longrightarrow
e^{-}+e^{-}+\nu+\overline{\nu}\hspace{1.75cm}
(electron-neutrino\hspace{0.2cm}bremsstrahlung),$$
$$(3)\hspace{0.5cm}e^{-}\longrightarrow^{magnetic\hspace{0.2cm}field}e^{-}+
\nu+\overline{\nu}\hspace{2.6cm}(neutrino\hspace{0.2cm}synchrotron\hspace{0.2cm}process),$$
$$(4)\hspace{0.5cm}e^{-}+Z\longrightarrow^{magnetic\hspace{0.2cm}field}
e^{-}+Z+\nu+\overline{\nu}\hspace{0.2cm}(bremsstrahlung\hspace{0.2cm}
in\hspace{0.2cm}magnetic field).$$ It merits careful attention
that the (3) and (4) of the above processes are considered in
presence of a strong magnetic field since the stellar core may be
magnetized in some cases. The presence of magnetic field may
influence some neutrino emission processes. In the subsequent
chapters we have considered all four processes and carried out a
detailed calculations in the framework of the electro-weak theory
with a minimal extension of the Standard Model as the presence of
small neutrino mass has been assumed. In the Standard Model
neutrino was considered a massless particle, but some
phenomenological consequences viz. `solar neutrino problem' and
`atmospheric neutrino anomaly' \cite{Santo} led to the concept of
tiny neutrino mass. In some works \cite{Dicus2000, Dodelson}
related to the photon-neutrino interaction process
($\gamma+\nu\longrightarrow \gamma+\nu$) this small neutrino mass
was already taken into account. In our calculations we have put
the neutrino mass by hand and since it is very small the basic
assumptions of the Standard Model are not violated. We have also
considered all three types of neutrino $(\nu_{e}, \nu_{\mu},
\nu_{\tau})$ in our entire calculations. With all such
considerations we have calculated exhaustively the four neutrino
emission processes. In the first two chapters the neutrino
emission processes in absence of magnetic field have been
considered, whereas in the next two we have studied the processes
in presence of strong magnetic fields. In our entire paper we
have followed the standard notations of QFT \cite{Greiner1} and
Standard Model \cite{Greiner2} and used the Gaussian system of
units.

\section{Photo-Coulomb Neutrino Process}
\subsection{Introduction}
In 1961 Matinyan and Tsilosani \cite{Matinyan} indicated that
there might exist a new kind of mechanism that would contribute to
the energy loss during a particular stage of stellar evolution.
Gell-Mann \cite{Gell-Mann} showed that the process wherein two
$\gamma$ quanta are converted into neutrino-antineutrino pair
$(\gamma+\gamma\longrightarrow\nu+\overline{\nu})$, is forbidden
under current current coupling theory. He argued that a local weak
interaction, to the lowest order of $G_{F}$ (Fermi coupling
constant), would lead to a vanishing amplitude for the above
process. If one of the photon is replaced by a coulomb field the
process would be no longer forbidden and termed as photo-coulomb
neutrino process $(\gamma+Z\longrightarrow Z+\nu+\overline{\nu})$.
Matinyan and Tsilosani \cite{Matinyan} calculated the scattering
cross-section and neutrino luminosity for this process in the
non-relativistic approximations. This calculation indicated that
at the density $\rho=10^{5}$ $gm/cc$ and temperature $5\times
10^{8}$ K the energy loss rate via this process would be $10^{3}$
$erg/gm-sec$, which is much greater than the energy loss rate due
to the photon emissions at the same temperature and density. They
pointed out that the rate of energy release obtained by the
photo-coulomb neutrino process would be many times smaller than
that in hydrogen reactions, and therefore the neutrino loss can
compete with the thermonuclear loss provided the thermonuclear
reactions in the star are non-existent and the star is
characterized by a large value of $Z$ (nucleus). They compared
their result (neutrino luminosity) with that of the neutrino
bremsstrahlung process obtained by Gandel'man and Pinaev
\cite{Gandel'man}. Their investigation showed that in a wide range
of stellar density and temperatures this mechanism might be
dominant over the neutrino bremsstrahlung. In 1962 Rosenberg
\cite{Rosenberg} considered this process in detail according to
the current current coupling theory, assuming the existence of a
direct electron-neutrino weak interaction. Chiu and Morrison
\cite{Chiu1960} suggested in their paper that the direct
electron-neutrino interaction implied by the current current
coupling theory would give rise to the processes that might
increase the neutrino luminosity of stars, thereby affecting
stellar evolution. The matrix element of the photo-coulomb
neutrino mechanism contains a diverging term arising out of the
presence of electronic loop in the Feynman diagrams. Rosenberg
imposed the gauge invariant condition explicitly to obtain a
finite result which, he indicated, would be analogous to the
situation in standard quantum electrodynamics leading to Ward's
identity. Neutrino was considered massless in his calculations.
According to his considerations only electron type of
neutrino-antineutrino pair would be emitted in this process. This
had rationale as the process was studied in the direct
electron-neutrino weak interaction theory. Rosenberg calculated
the scattering cross-section and the energy loss rate for the low
photon energy. He argued that for $T\geq 10^{9}$ K the
photo-coulomb neutrino would give smaller contribution in
comparison to pair annihilation or photo-neutrino process. In 1969
Raychaudhuri \cite{Raychaudhuri1} considered this process assuming
that the photons could interact weakly with the neutrinos. He
calculated the scattering cross-section and energy loss rate in
the framework of this photon neutrino weak coupling theory. He
also computed the neutrino luminosity expressed in the unit of
solar luminosity and compared the result with that obtained in the
current current coupling theory. It was observed from the
comparison that at the density $\rho=10^{5}$ $gm/cc$ and in the
temperature range $T\geq 10^{8}$ K, the photo-coulomb neutrino
calculated in the current current coupling theory would be
dominating over that in the photon neutrino weak coupling theory.
This showed that current current coupling theory could explain
this process successfully and thus strongly supported Rosenberg's
work on photo-coulomb neutrino process.\\\indent The above works
motivated us to study this process in the framework of
electro-weak theory. It is known that in the low energy region the
electro-weak theory conforms to the current current coupling
theory and since the photo-coulomb neutrino process was
successfully explained by current current coupling theory, it
should give, at least in principle, the same result when
calculating in the electro-weak theory. According to the
electro-weak theory the neutrino pair emission always takes place
through the exchange of intermediate vector boson and the
interaction Lagrangian contains neutral current as well as charged
current. Therefore, we consider the diagram exhibiting
$e^{-}-W^{-}-\nu_{e}$ effect, which has a significant contribution
in calculating the scattering cross-section. In addition to that,
all three types of neutrino are involved in this process, which
contribute a greater cross-section compared to what was calculated
by Rosenberg \cite{Rosenberg}. It is worth noting that there is no
scope to consider muon and tau neutrinos under the local
electron-neutrino interaction theory. We also like to consider the
process in a more generalized manner and hence the virtual
electronic loop is replaced by a fermionic loop consisting of all
first generation fermions, i.e., electrons along with $u$ and $d$
quarks. The consideration is quite reasonable as it includes all
possible fermions in the theory. All possible Feynman diagrams are
taken into account, though many of them would have almost
negligible effect in the matrix element. We are going to calculate
the total scattering cross-section and to obtain an analytical
expression for the energy loss rate depending on absolute
temperature. We also like to compute the neutrino luminosity and
the process is studied in the temperature range $10^{8}-10^{9}$ K,
which characterizes a low energy region. The calculation is also
carried out when the energy of the incoming photon is quite high
compared to the rest energy of the electron (but still very much
smaller than that of intermediate boson). Though this case is not
much relevant in the astrophysical scenario, we should give more
emphasis on the process wherein low energy photon is involved. An
important point to be noted that is unlike Rosenberg's
consideration \cite{Rosenberg} we consider the neutrino mass. The
nucleus involved in this process is taken to be at rest. In
principle, it may not be at rest, but because of the heavy rest
mass its motion is ignored. The situation is quite similar to the
process like coulomb scattering in quantum electrodynamics. The
calculations related to the photo-coulomb neutrino process are
shown here in more detail. \footnote {This result is published in
{\it Astroparticle physics} \cite{Bhattacharyya1}.}

\subsection{Calculation of scattering cross-section}
In the photo-coulomb neutrino process a real photon interacts with
a space-like photon coming out of a heavy nucleus, which is
assumed to be at rest. All Feynman diagrams pertaining to this
process are shown in Figures-1, 2 and 3, though the diagrams
in Figure-3 would have negligible contribution as each of those
contains more than one intermediate bosons. According to
electro-weak theory the photon cannot directly interact with
neutrino; therefore, the diagram has a loop. This loop gives
unwanted divergent term in the matrix element. Taking Figure-1
and Figure-2 the matrix element can be constructed as follows:
$$ M_{fi}=\frac{ie^{2}}{(2\pi)^{4}}A_{0}(\overrightarrow{k_{2}})
\varepsilon_{1\sigma}\varepsilon_{2\rho}J_{\mu}[\frac{4g_{Z}^{2}}{M_{Z}^{2}}
\sum_{f}I_{3}(f)C_{f}q_{f}^{2}R_{f}^{\sigma\rho\mu}+\frac{g_{W}^{2}}{M_{W}^{2}}
R_{e}^{\sigma\rho\mu}]\eqno{(1.2.1)}$$ where,
$$ R_{f}^{\sigma\rho\mu}(k_{1},k_{2})=2\int
Tr[\frac{m_{f}+q_{\alpha}\gamma^{\alpha}+k_{1\alpha}\gamma^{\alpha}}{(q+k_{1})^{2}-m_{f}^{2}
+i\epsilon}\gamma^{\sigma}\frac{m_{f}+q_{\alpha}\gamma^{\alpha}}{q^{2}-m_{f}^{2}
+i\epsilon}\gamma^{\rho}\frac{m_{f}+q_{\alpha}\gamma^{\alpha}
-k_{2\alpha}\gamma^{\alpha}}{(q-k_{2})^{2}-m_{f}^{2}+i\epsilon}\gamma^{\mu}\gamma_{5}]
d^{4}q\eqno{(1.2.2)}$$
$$J_{\mu}=\overline{u}(p_{1})\gamma_{\mu}(1-\gamma_{5})v(p_{2})\eqno{(1.2.3)}$$
$$\varepsilon_{2\rho}=\delta_{\rho0}\eqno{(1.2.4a)}$$
$$A_{0}(\overrightarrow{k_{2}})=-\frac{4\pi Ze}{\mid
\overrightarrow{k_{2}}\mid^{2}}\eqno{(1.2.4b)}$$ $C_{f}$ is the
color factor and $q_{f}$ is the fermionic charge taken in the unit
of $\mid e \mid$; $f$ stands for all first generation quarks and
electron. In the equation (1.2.1) $g_{Z}$ and $g_{W}$ stand for
the coupling constants for intermediate $Z$ and $W$-bosons,
respectively.
\\\indent The expression
$\varepsilon_{1\sigma}\varepsilon_{2\rho}R_{f}^{\sigma\rho\mu}$ is
symmetric under the interchange of the labels 1 and 2. An
important part of this calculation is to evaluate the integral
$R_{f}^{\sigma\rho\mu}(k_{1},k_{2})$; or in other words, to remove
the divergent term arising in it. Rosenberg \cite{Rosenberg}
imposed the gauge invariant condition explicitly to solve this
problem. We have used almost the same procedure and it is shown in
detail in the Appendix at the end of this chapter. After removing
the divergent factor we can obtain the following expression.
$$\varepsilon_{1\sigma}\varepsilon_{2\rho}J_{\mu}R^{\sigma\rho\mu}_{f}=
\mid\overrightarrow{k_{2}}\mid^{2}I_{f}(k_{1},k_{2})\overrightarrow{J}.
(\overrightarrow{\varepsilon_{1}}\times\overrightarrow{k_{1}})\eqno{(1.2.5)}$$
where,
$$ I_{f}(k_{1},k_{2})=16\pi^{2}\int_{0}^{1}dx_{1}\int_{0}^{1-x_{1}}\frac{x_{1}
(x_{1}+x_{2}-1)}{x_{1}(1-x_{1})k^{2}_{2}+2x_{1}x_{2}(k_{1}k_{2})-m_{f}^{2}}dx_{2}\eqno{(1.2.6)}$$
We can use the equation (1.2.5) to get the expression for
$M_{fi}$. Our aim is to calculate the scattering cross-section for
this process. The scattering cross-section for the photo-coulomb
neutrino process is calculated by using the following formula.
$$\sigma=\frac{4\pi}{2k_{1}^{0}}\int\frac{\sum \mid M_{fi}\mid
^{2}}{2p_{1}^{0}2p_{2}^{0}}2\pi
\delta(p_{1}^{0}+p_{2}^{0}-k_{1}^{0})\frac{d^{3}p_{1}}{(2\pi)^{3}}
\frac{d^{3}p_{2}}{(2\pi)^{3}}\eqno{(1.2.7)}$$ Here, the summation
is taken over the spin states of the neutrino and antineutrino. So
our task is to calculate the term $\sum \mid M_{fi}\mid ^{2}$. For
that purpose we calculate the term
$\sum\mid\overrightarrow{J}.(\overrightarrow{\varepsilon_{1}}
\times\overrightarrow{k_{1}})\mid^{2}$ and obtain
$$\sum\mid\overrightarrow{J}.(\overrightarrow{\varepsilon_{1}}
\times\overrightarrow{k_{1}})\mid^{2}=\frac{2}{m_{\nu_{e}}^{2}}
(k_{1}^{0})^{2}[(p_{1}^{0})^{2}+\mid\overrightarrow{p_{1}}\mid^{2}\cos
2\theta]\eqno{(1.2.8)}$$ [$\theta$ is the angle between the
directions of motion of the incoming photon and the neutrino]\\
In the low energy limit the term $ I_{f}(k_{1},k_{2})$ is
approximated to $I_{f}(0,0)$, since $k^{0}_{1}\ll m_{e}$. Thus we
can obtain the expression of $ \sum \mid M_{fi}\mid^{2}$. To
calculate the scattering cross-section we carry out the
integration given in the equation (1.2.7) and obtain the
following.
$$\int\frac{\sum \mid M_{fi}\mid
^{2}}{2p_{1}^{0}2p_{2}^{0}}2\pi
\delta(p_{1}^{0}+p_{2}^{0}-k_{1}^{0})\frac{d^{3}p_{1}}{(2\pi)^{3}}
\frac{d^{3}p_{2}}{(2\pi)^{3}}$$$$=\frac{Z^{2}\alpha^{3}}{4(2\pi)^{9}}
\frac{(k_{1}^{0})^{7}}{m_{\nu_{e}}^{2}}[\frac{4g_{Z}^{2}}{M_{Z}^{2}}
\sum_{f}I_{3}(f)C_{f}q_{f}^{2}I_{f}+\frac{g_{W}^{2}}{M_{W}^{2}}I_{e}]^{2}
[1-10(\frac{m_{\nu_{e}}}{k_{1}^{0}})^{2}+16(\frac{m_{\nu_{e}}}{k_{1}^{0}})^{4}]\eqno{(1.2.9)}$$
Finally, introducing some normalized factors and using equation
(1.2.8) with some simplifications the scattering cross-section is
obtained as follows.
$$\sigma_{\nu_{e}}=4.92\times10^{-28}\times Z^{2}(\frac{k_{1}^{0}}{m_{e}})^{6}
[1+\frac{m_{e}^{2}}{6m_{u}^{2}}-\frac{2m_{e}^{2}}{3m_{d}^{2}}]^{2}
[1-10(\frac{m_{\nu_{e}}}{k_{1}^{0}})^{2}+16(\frac{m_{\nu_{e}}}{k_{1}^{0}})^{4}]
\hspace{0.2cm} cm^{2}\eqno{(1.2.10)}$$ This is the scattering
cross-section obtained for the emission of electron type of
neutrino. The other two types of neutrino may be emitted by this
process and the calculations for obtaining scattering
cross-section will be almost same as the previous case , though
there will be no contribution for $W$-boson exchange. This part of
scattering-cross section is calculated as
$$\sigma_{\nu_{\mu},\nu_{\tau}}=2.49\times10^{-28}\times Z^{2}
(\frac{k_{1}^{0}}{m_{e}})^{6}[1+\frac{m_{e}^{2}}{3m_{u}^{2}}-
\frac{4m_{e}^{2}}{3m_{d}^{2}}]^{2}$$$$[2-10(\frac{m_{\nu_{\mu}}}{k_{1}^{0}})^{2}-10
(\frac{m_{\nu_{\tau}}}{k_{1}^{0}})^{2}+16(\frac{m_{\nu_{\mu}}}{k_{1}^{0}})^{4}
+16(\frac{m_{\nu_{\tau}}}{k_{1}^{0}})^{4}]\hspace{0.5cm}
cm^{2}\eqno{(1.2.11)}$$ Considering all three types of neutrino
the total scattering cross-section will be
$$\sigma=\sigma_{\nu_{e}}+\sigma_{\nu_{\mu},\nu_{\tau}}\eqno{(1.2.12)}$$
whereas $\sigma_{\nu_{e}}$ and $\sigma_{\nu_{\mu},\nu_{\tau}}$ are
given by the equations (1.2.10) and (1.2.11). Taking the leading
term resulting from the approximations we get
$$\sigma\approx 2.87\times10^{-55}\times Z^{2}(\frac{E}{m_{e}c^{2}})^{6}
\hspace{0.5cm} cm^{2}\eqno{(1.2.13)}$$ where,
$$E=k_{1}^{0}c\eqno{(1.2.14)}$$
The scattering cross-section, given by the equation (1.2.13), is
approximately six times of the scattering cross-section calculated
by Rosenberg \cite{Rosenberg}.
 \\\indent Let us consider the case when the
energy is much higher, i.e.,
$$ M_{Z}^{2}> M_{W}^{2}\gg
(p_{1}+p_{2})^{2}=(k_{1}+k_{2})^{2}\gg m_{f}^{2}$$ In this case
the term $ I_{f}(k_{1},k_{2})$ will develop an imaginary part as
$$I_{f}(k_{1},k_{2})\approx\frac{2\pi^{2}}{(k_{1}^{0})^{2}}[2-i\pi]\eqno{(1.2.15)}$$
and the scattering cross-section is calculated as
$$\sigma_{high}\approx 6.64\times 10^{-55}\times Z^{2}(\frac{E}{m_{e}c^{2}})^{2}
\hspace{0.5cm} cm^{2}\eqno{(1.2.16)}$$ This high energy case has
no significance in the stellar energy loss.
\subsection{Calculation of energy loss rate and luminosity}
The energy loss rate is calculated to see how much energy can be
radiated from the stellar core per unit  mass per unit time as the
result of photo-coulomb neutrino process. We can calculate the
energy loss rate with the aid of the following formula.
$$ \rho\mathcal{E}_{\nu}=2\sum_{i}N_{i}Z_{i}^{2}\int\frac{d^{3}k}{(2\pi)^{3}}
[e^{\frac{\hbar\omega}{\kappa T}}-1]^{-1}(\hbar\omega
c)\sigma\eqno{(1.3.1)}$$ with $\hbar\omega=\hbar
\mid\overrightarrow{k}\mid c=$ photon energy, $N_{i}=$ number of
nuclei of atomic number $ Z_{i}$ per unit volume (cubic centimeter)
and $\rho=$ stellar density\\
In the low energy limit and at the fixed density $\rho=10^{5}$
$gm/cc$ the energy loss rate is obtained as
$$\mathcal{E}_{\nu}=0. 99 \times 10^{-9} T_{8}^{10} \hspace{0.2 cm} erg/gm-sec
\eqno{(1.3.2)}$$ where, $ T_{8}= T\times 10^{-8}$ \\We can compare
this with the energy loss rate obtained by Rosenberg
\cite{Rosenberg} according to the current current coupling theory.
\\\indent We are now going to calculate the neutrino luminosity expressed
in the unit of solar luminosity in the non-degenerate stellar
core. It is calculated from the following formula \cite{Hayashi}
$$\frac{L_{\nu}}{L_{\odot}}=\frac{0.736}{\mu_{e}^{2}}A^{\frac{3}{2}}
(\frac{1-\beta}{\beta})^{\frac{1}{2}}\mathcal{E}_{\nu}\eqno{(1.3.3)}$$
where,
$$\mathcal{E}_{\nu}=\mathcal{E}_{\nu}^{0}\rho^{p}T^{s}$$
and
$$A=\frac{3n+3}{[n-3(1-\beta)(1+p)+\beta s]}$$
We see that $\beta$ plays an important role to find the neutrino
luminosity. In the stellar body the pressure $P$ is composed of
two parts: radiation pressure $P_{r}$ and gas pressure $P_{g}$.
The fraction of the gas pressure is denoted by $\beta$, i.e.,
$$ P_{g}=\beta P$$
In the case of perfect gas, the gas pressure is given by
$$ P_{g}=\frac{\kappa \rho T}{\mu_{H} m_{H}}\eqno{(1.3.4)}$$
where $ \mu_{H}$ is the mean molecular weight. In the
non-degenerate gas we cannot neglect radiation pressure and it
becomes
$$P_{r}=\frac{aT^{4}}{3}=(1-\beta)P\eqno{(1.3.5)}$$
where $a$ represents Steafan-Boltzman constant.
\\Finally, the neutrino luminosity for the photo-coulomb neutrino process is obtained as
$$ \frac{L_{\nu}}{L_{\odot}}=1.62 \times
10^{-10}\frac{(1-\beta)^{\frac{1}{2}}}{\beta^{2}}T_{8}^{10}\eqno{(1.3.6)}$$
\\In the similar manner the neutrino luminosity can be obtained from Rosenberg's result.
In the Table-1 we compare these two results in the temperature
range $10^{8}-10^{9}$ K  and it shows that the result obtained in
electro-weak theory dominates that obtained in current current
coupling theory. The table also shows that both the results seem
to be significant near the temperature $10^{9}$ K.
\subsection{Discussion}
Photo-coulomb neutrino process is studied anticipating that it may
play an important role in a particular stage during stellar
evolution. This process is very much similar to the neutrino
bremsstrahlung process. In the bremsstrahlung the neutrino pair
emission takes place when an electron collides with a heavy
nucleus, whereas in the photo-coulomb neutrino process the photon
interacts with nucleus to emit neutrino-antineutrino pair. It has
already been mentioned that earlier this process was calculated
and explained by Rosenberg. However, he did not design a general
structure of this kind of interaction in the framework of
electro-weak theory. To realize a complete picture we have
intended to consider this process according to the Standard Model.
In the framework of electro-weak theory we have calculated the
scattering cross-section which is larger than that obtained by
Rosenberg. The scattering cross-section is obtained in the low
energy limit, that is, when the energy of the incoming photon is
much lower than the rest energy of the electron. This situation
can be realized in terms of astrophysical observable quantities
$-$ the temperature and density. The energy being much lower than
the rest energy of electron signifies that the stellar temperature
(core temperature) is well below $5.93\times 10^{9}$ K and density
is less than or equal to $10^{5}$ $gm/cc$. For that reason we have
computed the neutrino luminosity in the temperature range
$10^{8}-10^{9}$ K and at the density $10^{5}$ $gm/cc$. It has also
been found from Table-1 that the relative luminosity obtained in
the electro-weak theory is greater than that obtained in the
current current coupling theory. If we study this table carefully,
we see that at the beginning of the table the relative luminosity
is very small, but as we move towards the end of the table, i.e.,
towards the temperature $10^{9}$ K, the relative luminosity
increases rapidly and the process exhibits the significant effect.
In this temperature range the neutrino luminosity is comparable to
the photon luminosity, which clearly indicates that the
photo-coulomb neutrino process is important in the temperature
range indicated in the Table-1.\\\indent Let us now consider the
case when the energy of the photon is much greater than
$m_{f}c^{2}$ (rest energy of any kind of first generation
fundamental fermion), but well below the rest energy of the
intermediate bosons. Astrophysically this is not an interesting
case. The neutrino luminosity for the photo-coulomb neutrino
process is negligibly small in this energy range. Therefore, the
process has no significant role in high temperature and density.
In particular the process has the maximum effect at the
temperature close to $10^{9}$ K and density $10^{5}$ $gm/cc$. Thus
the photo-coulomb neutrino process plays a crucial role for energy
loss in the star having low luminosity.
\subsection{Appendix}
The term $R^{\sigma\rho\mu}(k_{1},k_{2})$ related to the fermionic
loop can be expressed as
$$R^{\sigma\rho\mu}(k_{1},k_{2})=A_{1}k_{1\tau}\epsilon^{\tau\sigma\rho\mu}+
A_{2}k_{2\tau}\epsilon^{\tau\sigma\rho\mu}+A_{3}k_{1}^{\rho}k_{1\xi}k_{2\tau}
\epsilon^{\xi\tau\sigma\mu}+$$$$
A_{4}k_{2}^{\rho}k_{1\xi}k_{2\tau}\epsilon^{\xi\tau\sigma\mu}+
A_{5}k_{1}^{\sigma}k_{1\xi}k_{2\tau}\epsilon^{\xi\tau\rho\mu}
+A_{6}k_{2}^{\sigma}k_{1\xi}k_{2\tau}
\epsilon^{\xi\tau\rho\mu}\eqno{(1)}$$ where,
$$A_{3}(k_{1},k_{2})=-I_{11}(k_{1},k_{2})=-A_{6}(k_{1},k_{2})$$
$$A_{4}(k_{1},k_{2})=[I_{20}(k_{1},k_{2})-I_{10}(k_{1},k_{2})]
=-A_{5}(k_{1},k_{2})\eqno{(2)}$$
$$I_{st}(k_{1},k_{2})=16\pi^{2}\int_{0}^{1}dx_{1}\int_{0}^{1-x_{1}}
\frac{x_{1}^{s}x_{2}^{t}}{x_{2}(1-x_{2})k^{2}_{1}+x_{1}(1-x_{1})
k^{2}_{2}+2x_{1}x_{2}(k_{1}k_{2})-m^{2}}dx_{2}\eqno{(3)}$$
[Instead of $R^{\sigma\rho\mu}_{f}$ we present here
$R^{\sigma\rho\mu}$.]\\The terms $A_{1}$ and $A_{2}$ are
represented by the divergent integrals, which create a difficulty.
To get rid of this difficulty the following conditions (gauge
invariancy) are imposed.
$$k_{1\sigma}R^{\sigma\rho\mu}=k_{2\rho}R^{\sigma\rho\mu}=0\eqno{(4)}$$
which give
$$[-A_{2}+k_{1}^{2}A_{5}+(k_{1}k_{2})A_{6}]k_{1\xi}
k_{2\tau}\epsilon^{\xi\tau\rho\mu}=0$$
$$[-A_{1}+k_{2}^{2}A_{4}+(k_{1}k_{2})A_{3}]k_{1\xi}
k_{2\tau}\epsilon^{\xi\tau\sigma\mu}=0\eqno{(5)}$$
 Let us take
$$A_{2}=k_{1}^{2}A_{5}+(k_{1}k_{2})A_{6}\eqno{(6)}$$
$$A_{1}=k_{2}^{2}A_{4}+(k_{1}k_{2})A_{3}\eqno{(7)}$$
and use the following identity.
$$(af)\mid bcde\mid+(bf)\mid cdea\mid+(cf)\mid deab\mid+(df)\mid eabc\mid+(ef)
\mid abcd\mid=0\eqno{(8)}$$ where,
$$\mid abcd\mid=a_{\alpha}b_{\beta}c_{\gamma}d_{\delta}\epsilon^{\alpha\beta\gamma\delta}
\eqno{(9)}$$ Inserting the expressions of $A_{1}$ and $A_{2}$ into
the equation (1) along with using the identity (8) and the
conditions
$$(\varepsilon_{1}k_{1})=(\varepsilon_{2}k_{2})=0\eqno{(10)}$$
it can be obtained
$$\varepsilon_{1\sigma}\varepsilon_{2\rho}J_{\mu}R^{\sigma\rho\mu}=
k_{1}^{2}(A_{3}+A_{5})\mid
k_{2}J\varepsilon_{1}\varepsilon_{2}\mid+k_{2}^{2}(A_{4}+A_{6})
\mid
k_{1}J\varepsilon_{1}\varepsilon_{2}\mid$$$$+[A_{3}(k_{1}J)-A_{6}(k_{2}J)]\mid
k_{1}k_{2}\varepsilon_{1}\varepsilon_{2}\mid\eqno{(11)}$$ The
above expression can be simplified by expressing $A_{5}$ and
$A_{6}$ in terms of $A_{4}$ and $A_{3}$ respectively from the
conditions given by (2). Again, it is known that $k_{1}$
represents the four momentum of the external photon and thus the
following condition holds.
$$k_{1}^{2}=0\eqno{(12)}$$
Let us choose the frame of reference such that
$$\overrightarrow{k_{1}}+\overrightarrow{k_{2}}=
\overrightarrow{p_{1}}+\overrightarrow{p_{2}}=0\eqno{(13)}$$ In
this frame it is found that
$$\mid k_{1}k_{2}\varepsilon_{1}\varepsilon_{2}\mid=\overrightarrow{k_{2}}.
(\overrightarrow{\varepsilon_{1}}\times\overrightarrow{k_{1}})=0\eqno{(14)}$$
Therefore, finally the equation (11) is written as
$$\varepsilon_{1\sigma}\varepsilon_{2\rho}J_{\mu}R^{\sigma\rho\mu}=k_{2}^{2}(A_{4}-A_{3})
\mid k_{1}J\varepsilon_{1}\varepsilon_{2}\mid \eqno{(15)}$$ This
gives the equation (1.2.5).

\section{Electron-Neutrino Bremsstrahlung Process}
\subsection{Introduction}
In this chapter we are going to study the electron-neutrino
bremsstrahlung process in which two electrons interact with each
other to produce neutrino antineutrino pair and thus in the final
state there are four fermionic particles (two electrons and a pair
of neutrino antineutrino). Here the interaction takes place
between two identical electrons. Through this interaction the
neutrino antineutrino pair is produced. In quantum electrodynamics
there is a similar kind of process, where photon is emitted
instead of neutrino antineutrino pair and whose scattering
cross-section was obtained by Wheeler and Lamb \cite{Wheeler}, and
the calculations were also carried out by I. Hodes \cite{Hodes}
and E. Huang \cite{Huang}. Previously, the electron-neutrino
bremsstrahlung process was considered by Cazzola and Saggion
\cite{Cazzola}. They calculated the energy loss rate in the
non-degenerate stellar region and discussed the implications of
this process. They took all possible diagrams to carry out the
calculations numerically by using Monte Carlo method, although
they did not obtain the scattering cross-section explicitly. They
considered only the non-degenerate case, though many important
stellar objects in the later phases such as white dwarves, neutron
stars etc. are degenerate in nature; therefore, the possibility of
occurring the electron-neutrino bremsstrahlung in degenerate stars
cannot be ruled out. Although they did not carry out the
calculations in the degenerate region, but they pointed the
process might be significant in the degenerate region as well. By
the work of Cazzola and Saggion \cite{Cazzola} we are motivated to
calculate the process in the framework of electro-weak theory. In
our work we consider both non-degenerate and degenerate cases
separately and discuss all possible outcomes of the
electron-neutrino bremsstrahlung process. We also visualize a
picture of the stellar regions in which the process will have some
significant effect. It cannot be denied that due to some
approximations a little bit deviation may occur from the original
result, but that will not deter to realize the physical picture.
The role of this process is studied thoroughly \footnote {This
result is published in {\it Journal of Physics G}
\cite{Bhattacharyya2}.} at different temperature and density
ranges during the late stages of the stellar evolution.
\subsection{Calculation of scattering cross-section}
In the electron-neutrino bremsstrahlung process a slight
complication arises since the identical particles (electrons) are
involved. It is not possible to identify which of the two outgoing
particles is the `target' particle for a particular `incident'
electron. In classical physics such identification can be done by
tracing out the trajectories. In quantum physics the two
alternatives are completely indistinguishable; therefore, the two
cases may interfere. There exist 8 possible Feynman diagrams shown
in Figure-4 and Figure-5. The total scattering amplitude for
all possible diagrams can be constructed as follows:
$$\mathcal{M}^{Z}=-\frac{4\pi i e^{2}g^{2}}{8\cos^{2}\theta_{W}
M_{Z}^{2}}[(\mathcal{M}_{1}^{Z}+\mathcal{M}_{2}^{Z}+\mathcal{M}_{3}^{Z}+
\mathcal{M}_{4}^{Z})-(\mathcal{M}_{5}^{Z}+\mathcal{M}_{6}^{Z}+\mathcal{M}_{7}^{Z}+
\mathcal{M}_{8}^{Z})]\eqno{(2.2.1)}$$
$$\mathcal{M}_{1}^{Z}=[\overline{u}(p'_{1})(C_{V}-C_{A}\gamma_{5})\gamma_{\rho}
\frac{(q^{\tau}\gamma_{\tau}+p_{1}^{'\tau}\gamma_{\tau}+m_{e})}
{(q+p'_{1})^{2}-m_{e}^{2}+i\epsilon}\gamma_{\mu}u(p_{1})]$$$$\hspace{4cm}
[\overline{u}(p'_{2})\frac{\gamma^{\mu}}{(p_{2}-p'_{2})^{2}+i\epsilon}u(p_{2})]
[\overline{u}_{\nu}(q_{1})(1-\gamma_{5})\gamma^{\rho}v_{\nu}(q_{2})]\eqno{(2.2.2)}$$
$$\mathcal{M}_{2}^{Z}=[\overline{u}(p'_{1})\gamma_{\mu}
\frac{(-q^{\tau}\gamma_{\tau}+p_{1}^{\tau}\gamma_{\tau}+m_{e})}
{(q-p_{1})^{2}-m_{e}^{2}+i\epsilon}(C_{V}-C_{A}\gamma_{5})\gamma_{\rho}u(p_{1})]$$$$
\hspace{4cm}[\overline{u}(p'_{2})\frac{\gamma^{\mu}}{(p_{2}-p'_{2})^{2}+i\epsilon}u(p_{2})]
[\overline{u}_{\nu}(q_{1})(1-\gamma_{5})\gamma^{\rho}v_{\nu}(q_{2})]\eqno{(2.2.3)}$$
$$\mathcal{M}_{3}^{Z}=\mathcal{M}_{1}^{Z}(p_{1}\leftrightarrow p_{2},p'_{1}\leftrightarrow
p'_{2})\hspace{2cm}\mathcal{M}_{4}^{Z}=\mathcal{M}_{2}^{Z}(p_{1}\leftrightarrow
p_{2},p'_{1}\leftrightarrow p'_{2})\eqno{(2.2.4)}$$
$$\mathcal{M}_{5}^{Z}=\mathcal{M}_{1}^{Z}(p'_{1}\leftrightarrow p'_{2})\hspace{2cm}
\mathcal{M}_{6}^{Z}=\mathcal{M}_{2}^{Z}(p'_{1}\leftrightarrow
p'_{2})$$$$
\mathcal{M}_{7}^{Z}=\mathcal{M}_{3}^{Z}(p'_{1}\leftrightarrow
p'_{2})\hspace{2cm}\mathcal{M}_{8}^{Z}=\mathcal{M}_{4}^{Z}(p'_{1}\leftrightarrow
p'_{2})\hspace{1cm}\eqno{(2.2.5)}$$ where,
$$C_{V}=-\frac{1}{2}+2\sin^{2}\theta_{W}\hspace{2cm} C_{A}=-\frac{1}{2}$$
The superscript $Z$ associated with the matrix element and each of
its component indicates that the neutrino antineutrino pair
emission takes place through the exchange of $Z$ boson.\\\indent
We have carried out our calculations in the CM frame in which
$$\overrightarrow{p_{1}}+\overrightarrow{p_{2}}
=\overrightarrow{p}'_{1}+\overrightarrow{p}'_{2}+
\overrightarrow{q_{1}}+\overrightarrow{q_{2}}=0\eqno{(2.2.6)}$$
where $\overrightarrow{q_{1}}$ and $\overrightarrow{q_{2}}$ are
the linear momenta of the neutrino and antineutrino, respectively.
In this frame we have to calculate the term
$|\mathcal{M}^{Z}|^{2}$ over the spin sum. It is not a very easy
task and can be done with some choices and approximations. Let us
first try to calculate the term $\sum |\mathcal{M}_{1}^{Z}|^{2}$
that can be written as
$$\sum |\mathcal{M}_{1}^{Z}|^{2}=X_{\rho\sigma\alpha\beta}(p_{1},p_{2},p'_{1},p'_{2})
Y^{\alpha\beta}(p_{2},p'_{2})N^{\rho\sigma}(q_{1},q_{2})\eqno{(2.2.7)}$$
$$X_{\rho\sigma\alpha\beta}(p_{1},p_{2},p'_{1},p'_{2})=\frac{1}{4m_{e}^{2}
|(q+p'_{1})^{2}-m_{e}^{2}+i\epsilon|^{2}}[C_{V}^{2}T_{1}-C_{A}^{2}T_{2}
+C_{V}C_{A}T_{3}-C_{V}C_{A}T_{4}]\eqno{(2.2.8)}$$
$$T_{1}=Tr[(p_{1}^{\tau}\gamma_{\tau}+m_{e})
\gamma_{\alpha}(P^{\tau}\gamma_{\tau}+m_{e})\gamma_{\rho}(p_{1}'^{\tau}\gamma_{\tau}+m_{e})
\gamma_{\sigma}(P^{\tau}\gamma_{\tau}+m_{e})\gamma_{\beta}]\eqno{(2.2.8a)}$$
$$T_{2}=Tr[(p_{1}^{\tau}\gamma_{\tau}+m_{e})
\gamma_{\alpha}(P^{\tau}\gamma_{\tau}+m_{e})\gamma_{\rho}(-p_{1}'^{\tau}\gamma_{\tau}+m_{e})
\gamma_{\sigma}(P^{\tau}\gamma_{\tau}+m_{e})\gamma_{\beta}]\eqno{(2.2.8b)}$$
$$T_{3}=Tr[(p_{1}^{\tau}\gamma_{\tau}+m_{e})
\gamma_{\alpha}(P^{\tau}\gamma_{\tau}+m_{e})\gamma_{\rho}(p_{1}'^{\tau}\gamma_{\tau}+m_{e})
\gamma_{5}\gamma_{\sigma}(P^{\tau}\gamma_{\tau}+m_{e})\gamma_{\beta}]\eqno{(2.2.8c)}$$
$$T_{4}=Tr[(p_{1}^{\tau}\gamma_{\tau}+m_{e})
\gamma_{\alpha}(P^{\tau}\gamma_{\tau}+m_{e})\gamma_{\rho}\gamma_{5}(-p_{1}'^{\tau}
\gamma_{\tau}+m_{e})\gamma_{\sigma}(P^{\tau}\gamma_{\tau}+m_{e})\gamma_{\beta}]\eqno{(2.2.8d)}$$
$$Y^{\alpha\beta}(p_{2},
p'_{2})=\frac{1}{m_{e}^{2}|(p_{2}-p'_{2})^{2}+i\epsilon|^{2}}
[p_{2}^{\alpha}p_{2}^{'\beta}+p_{2}^{\beta}p_{2}^{'\alpha}+
\{(p_{2}p'_{2})-m_{e}^{2}\}g^{\alpha\beta}]\eqno{(2.2.9)}$$
$$N^{\rho\sigma}(q_{1}, q_{2})=\frac{2}{m_{\nu}^{2}}
[q_{1}^{\rho}q_{2}^{\sigma}+q_{1}^{\sigma}q_{2}^{\rho}-
(q_{1}q_{2})g^{\rho\sigma}+iq_{1\tau_{1}}q_{2\tau_{2}}
\epsilon^{\tau_{1}\tau_{2}\rho\sigma}]\eqno{(2.2.10)}$$
$$q=q_{1}+q_{2}=(p_{1}+p_{2})-(p'_{1}+p'_{2})\eqno{(2.2.11a)}$$
$$P=q+p'_{1}\eqno{(2.2.11b)}$$
We evaluate various trace terms ($T_{1}, T_{2}, T_{3}$ and
$T_{4}$), present in the above equations, by using the formula
deduced in the Appendix. Instead of calculating the term
$\sum|\mathcal{M}_{1}^{Z}|^{2}$ it is easy to calculate
$$\int\sum|\mathcal{M}_{1}^{Z}|^{2}\frac{d^{3}q_{1}}{2q_{1}^{0}}\frac{d^{3}q_{2}}{2q_{2}^{0}}
\delta^{4}(q-q_{1}-q_{2})$$ Let us consider
$$I^{\rho\sigma}(q)=\frac{2}{m_{\nu}^{2}}\int
[q_{1}^{\rho}q_{2}^{\sigma}+q_{1}^{\sigma}q_{2}^{\rho}-
(q_{1}q_{2})g^{\rho\sigma}+iq_{1\tau_{1}}q_{2\tau_{2}}
\epsilon^{\tau_{1}\tau_{2}\rho\sigma}]
\frac{d^{3}q_{1}}{2q_{1}^{0}}\frac{d^{3}q_{2}}{2q_{2}^{0}}
\delta^{4}(q-q_{1}-q_{2})$$
$$=\frac{1}{m_{\nu}^{2}}(Aq^{2}g^{\rho\sigma}+Bq^{\rho}q^{\sigma})\eqno{(2.2.12)}$$
It is to be remembered that the neutrino mass is very small
compared to the magnitude of its linear momentum. This is valid
throughout our calculations, even in the non-relativistic case.
Even if it is comparably nearer to the magnitude of the linear
momentum, no such neutrino-antineutrino pair will be emitted and
the process will become superfluous. This is very much consistent
with the Standard Model which is based on the concept of massless
neutrino. Thus taking $m_{\nu}\ll q^{0}$ we evaluate the integral
$I^{\rho\sigma}(q)$ and find the value of $A$ and $B$ as follows:
$$A=-B=-\frac{\pi}{3}$$
We also have,
$$\int\sum|\mathcal{M}_{1}^{Z}|^{2}\frac{d^{3}q_{1}}{2q_{1}^{0}}\frac{d^{3}q_{2}}{2q_{2}^{0}}
\delta^{4}(q-q_{1}-q_{2})=X_{\rho\sigma\alpha\beta}(p_{1},p_{2},p'_{1},p'_{2})
Y^{\alpha\beta}(p_{2},p'_{2})I^{\rho\sigma}(q)\eqno{(2.2.13)}$$ In
the same manner the term $\Sigma|\mathcal{M}_{2}^{Z}|^{2}$ is
evaluated to obtain an expression similar to that given by the set
of equations (2.2.7) to (2.2.11a); in that case $P$ is replaced by
$Q$, where
$$Q=p_{1}-q$$
Evaluating the various trace terms rigorously and simplifying
those expressions we obtain
$$\int\sum|\mathcal{M}^{Z}|^{2}\frac{d^{3}q_{1}}{2q_{1}^{0}}\frac{d^{3}q_{2}}{2q_{2}^{0}}
\delta^{4}(q-q_{1}-q_{2})=F(p_{1}, p_{2}, p'_{1},
p'_{2})\eqno{(2.2.14)}$$ The right hand side of this equation is a
scalar obtained by the various combinations of the scalar product
of initial and final momenta of the electrons. Thus $F$ becomes
the function of either energies or momenta of incoming and
outgoing electrons. The scattering cross-section for this process
is calculated by using the formula
$$\sigma=\frac{\mathcal{S}}{4\sqrt{(p_{1}p_{2})^{2}-m_{e}^{4}}}N_{p_{1}}N_{p_{2}}
\frac{1}{(2\pi)^{2}}\int
\frac{N_{p'_{1}}d^{3}p'_{1}}{2p_{1}^{'0}(2\pi)^{3}}
\frac{N_{p'_{2}}d^{3}p'_{2}}{2p_{2}^{'0}(2\pi)^{3}}
N_{q_{1}}N_{q_{2}}F(p_{1},p_{2},p'_{1},p'_{2})\eqno{(2.2.15)}$$
Here, all incoming and outgoing particles are spin-$\frac{1}{2}$
fermions. For that reason $N_{i}$
$(i=p_{1},p_{2},p'_{1},p'_{2},q_{1}, q_{2})$ is twice the mass of
the corresponding fermion. The square root term present in the
denominator of the equation (2.2.15) comes from the incoming flux
which is directly proportional to the relative velocity of the
incoming electrons and written in the Lorentz invariant way. As
the final state contains the fermionic particles there must be a
non-unit statistical degeneracy factor $\mathcal{S}$ given by
$$\mathcal{S}=\prod_{l}\frac{1}{g_{l}!}$$ if there are $g_{l}$
particles of the kind $l$ in the final state. This factor arises
since for $g_{l}$ identical final particles there are exactly
$g_{l}!$ possibilities of arranging those particles, but only one
such arrangement is measured experimentally. We Calculate the
expression of $F(p_{1},p_{2},p'_{1},p'_{2})$ in the equation
(2.2.15) and then integrating that expression the scattering
cross-section is obtained. We are interested to obtain a clear
analytical expression and so do not use any numerical technique.
Instead, with some special choice of approximations we calculate
the integral present in (2.2.15). Now, we proceed to evaluate the
integral $\int\frac{F}{p^{'0}_{2}} d^{3}p'_{2}$. For that purpose
we use an approximation $F(p_{1},p_{2},p'_{1},p'_{2})\approx
F(\mid\overrightarrow{p}'_{2}\mid=\mid\overrightarrow{p}'_{1}\mid,...)$
within the integral sign and obtain
$$\int\frac{F}{p^{'0}_{2}} d^{3}p'_{2}=\frac{4\pi}{3}
\frac{\mid\overrightarrow{p}'_{1}\mid^{3}}{p^{'0}_{1}}
F(\mid\overrightarrow{p}'_{2}\mid=\mid\overrightarrow{p}'_{1}\mid,...)+\epsilon$$
The error term $\epsilon$ arises for using the approximation
mentioned above. Here, neglecting this error term we can use the
following approximation.
$$\int \frac{F}{p^{'0}_{2}} d^{3}p'_{2}\approx\frac{4\pi}{3}
\frac{\mid\overrightarrow{p}'_{1}\mid^{3}}{p^{'0}_{1}}
F(\mid\overrightarrow{p}'_{2}\mid=
\mid\overrightarrow{p}'_{1}\mid,...)\eqno{(2.2.16)}$$ Next, we
integrate over $d^{3}p'_{1}$ without any more approximation and
obtain the following expression of the scattering cross-section:
$$\sigma\approx\frac{(C_{V}^{2}+C_{A}^{2})}{9\pi^{2}}(\frac{eg}{M_{Z}\cos\theta_{W}})^{4}
\frac{(p^{0})^{2}}{\sqrt{1-(\frac{m_{e}}{p^{0}})^{2}}}
[\ln(\frac{p^{0}}{m_{e}})+f(p^{0},r)]\eqno{(2.2.17)}$$ where,
$$f(p^{0},r)=\ln r
-[r-\frac{m_{e}}{p^{0}}][14-\frac{16}{(1+\frac{C_{V}^{2}}{C_{A}^{2}})}
(\frac{m_{e}}{p^{0}})^{2}+\frac{3(1+\frac{3C_{V}^{2}}{2C_{A}^{2}})}
{(1+\frac{C_{V}^{2}}{C_{A}^{2}})}(\frac{m_{e}}{p^{0}})^{4}]$$
$$+\frac{1}{2}[r^{2}-(\frac{m_{e}}{p^{0}})^{2}]
[31-\frac{12(1+\frac{27C_{V}^{2}}{24C_{A}^{2}})}
{(1+\frac{C_{V}^{2}}{C_{A}^{2}})}(\frac{m_{e}}{p^{0}})^{2}-3(\frac{m_{e}}{p^{0}})^{4}]$$
$$-\frac{2}{3}[r^{3}-(\frac{m_{e}}{p^{0}})^{3}][7-3(\frac{m_{e}}{p^{0}})^{2}]
+\frac{1}{4}[r^{4}-(\frac{m_{e}}{p^{0}})^{4}]$$
$$-3(\frac{m_{e}}{p^{0}})^{2}\ln(\frac{rp^{0}}{m_{e}})
[\frac{4(1+\frac{27C_{V}^{2}}{24C_{A}^{2}})}{(1+\frac{C_{V}^{2}}{C_{A}^{2}})}
-(\frac{m_{e}}{p^{0}})^{2}
-\frac{1}{(1+\frac{C_{V}^{2}}{C_{A}^{2}})}(\frac{m_{e}}{p^{0}})^{4}]$$
$$+3(\frac{m_{e}}{p^{0}})^{2}[\frac{p^{0}}{m_{e}}-\frac{1}{r}]
[2-\frac{(1+\frac{2C_{V}^{2}}{3C_{A}^{2}})}
{(1+\frac{C_{V}^{2}}{C_{A}^{2}})}(\frac{m_{e}}{p^{0}})^{2}]$$
$$-\frac{3}{2}(\frac{m_{e}}{p^{0}})^{4}[(\frac{p^{0}}{m_{e}})^{2}-(\frac{1}{r})^{2}]
[1-\frac{1}{(1+\frac{C_{V}^{2}}{C_{A}^{2}})}
(\frac{m_{e}}{p^{0}})^{2}]\eqno{(2.2.17a)}$$ and
$$\frac{m_{e}}{p^{0}}<r=\frac{max(p^{'0}_{1},p^{'0}_{2})}{p^{0}}<1\eqno{(2.2.17b)}$$
$p^{0}$ represents the CM energy, i.e.,
$$p_{1}^{0}=p_{2}^{0}=p^{0}$$ whereas $p^{'0}_{1}$ and $p^{'0}_{2}$ stand
for energies of the outgoing electrons.\\\indent It is to be noted
that all three types of neutrino are involved in this process. So
far we have used the technique which is applicable for both muon
and tau neutrino, but for electron type of neutrino other 8
Feynman diagrams having $e-W^{-}-\nu_{e}$ effect contribute. Four
of them are related to the direct process (Figure-6) and rest
four represent exchange diagrams (Figure-7). These extra
diagrams are to be considered only for the electron type of
neutrino emission. In that case the matrix element has to be
modified as
$$M_{fi}=\mathcal{M}^{Z}+\mathcal{M}^{W}\eqno{(2.2.18)}$$
where,
$$\mathcal{M}^{W}=-\frac{4\pi i e^{2}g^{2}}{8M_{W}^{2}}
[(\mathcal{M}_{1}^{W}+\mathcal{M}_{2}^{W}+\mathcal{M}_{3}^{W}+
\mathcal{M}_{4}^{W})-(\mathcal{M}_{5}^{W}+\mathcal{M}_{6}^{W}+\mathcal{M}_{7}^{W}+
\mathcal{M}_{8}^{W})]\eqno{(2.2.19)}$$
$$\mathcal{M}_{1}^{W}=[\overline{u}(p'_{1})(1-\gamma_{5})\gamma_{\rho}
\frac{(q^{\tau}\gamma_{\tau}+p_{1}^{'\tau}\gamma_{\tau}+m_{e})}
{(q+p'_{1})^{2}-m_{e}^{2}+i\epsilon}\gamma_{\mu}v_{\nu}(q_{2})]$$$$\hspace{4.2cm}
[\overline{u}(p'_{2})\frac{\gamma^{\mu}}{(p_{2}-p'_{2})^{2}+i\epsilon}u(p_{2})]
[\overline{u}_{\nu}(q_{1})(1-\gamma_{5})\gamma^{\rho}u(p_{1})]\eqno{(2.2.20a)}$$
$$\mathcal{M}_{2}^{W}=[\overline{u}(p'_{1})\gamma_{\mu}
\frac{(-q^{\tau}\gamma_{\tau}+p_{1}^{\tau}\gamma_{\tau}+m_{e})}
{(q-p_{1})^{2}-m_{e}^{2}+i\epsilon}(1-\gamma_{5})\gamma_{\rho}v_{\nu}(q_{2})]$$$$
\hspace{4.2cm}[\overline{u}(p'_{2})\frac{\gamma^{\mu}}{(p_{2}-p'_{2})^{2}+i\epsilon}u(p_{2})]
[\overline{u}_{\nu}(q_{1})(1-\gamma_{5})\gamma^{\rho}u(p_{1})]\eqno{(2.2.20b)}$$
Other $\mathcal{M}^{W}_{i}$'s (i=3,....8) have similar expressions
as defined in the equations (2.2.4) and (2.2.5). We use Fierz
rearrangement to obtain the full expression for $M_{fi}$
containing the contributions for both $Z$ and $W$ bosons exchanged
diagrams. If we introduce Fierz rearrangement on
$\mathcal{M}_{1}^{W}$ in $(2.2.20a)$ and add it to (2.2.2) we
obtain
$$\mathcal{M}_{1}=[\overline{u}(p'_{1})(C_{V}'-C_{A}'\gamma_{5})\gamma_{\rho}
\frac{(q^{\tau}\gamma_{\tau}+p_{1}^{'\tau}\gamma_{\tau}+m_{e})}
{(q+p'_{1})^{2}-m_{e}^{2}+i\epsilon}\gamma_{\mu}u(p_{1})]$$$$\hspace{4.2cm}
[\overline{u}(p'_{2})\frac{\gamma^{\mu}}{(p_{2}-p'_{2})^{2}+i\epsilon}u(p_{2})]
[\overline{u}_{\nu}(q_{1})(1-\gamma_{5})\gamma^{\rho}v_{\nu}(q_{2})]\eqno{(2.2.21a)}$$
where,$$C'_{V}=\frac{1}{2}+2\sin^{2}\theta_{W}\hspace{2cm}
C'_{A}=-\frac{1}{2}$$ Thus the complete scattering matrix takes
the form as
$$M_{fi}=-\frac{4\pi i e^{2}G_{F}}{\sqrt{2}}
[(\mathcal{M}_{1}+\mathcal{M}_{2}+\mathcal{M}_{3}+
\mathcal{M}_{4})-(\mathcal{M}_{5}+\mathcal{M}_{6}+\mathcal{M}_{7}+
\mathcal{M}_{8})]\eqno{(2.2.21b)}$$ We have
$$\frac{G_{F}}{\sqrt{2}}=\frac{g^{2}}{8M_{W}^{2}}=\frac{g^{2}}{8M_{Z}^{2}\cos^{2}\theta_{W}}$$
Each $\mathcal{M}_{i}$ (i=1,2.....8) is formed by adding
$\mathcal{M}^{Z}_{i}$ and $\mathcal{M}^{W}_{i}$ (after Fierz
rearrangement). Then we proceed in the same way as before and
calculate the scattering cross-section for the electron type of
neutrino as
$$\sigma_{\nu_{e}}\approx\frac{4(C_{V}^{'2}+C_{A}^{'2})}{9\pi^{2}}\alpha^{2}G_{F}^{2}
\frac{(p^{0})^{2}}{\sqrt{1-(\frac{m_{e}}{p^{0}})^{2}}}
[\ln(\frac{p^{0}}{m_{e}})+f_{\nu_{e}}(p^{0},r)]\eqno{(2.2.22)}$$
The expression of $f_{\nu_{e}}(p^{0},r)$ present in the equation
(2.2.22) is almost similar to the expression of $f(p^{0},r)$ given
by the equation (2.2.17a). In fact when we replace $C_{V}$ and
$C_{A}$ present in $f(p^{0},r)$ by $C'_{V}$ and $C'_{A}$,
respectively, the expression $f_{\nu_{e}}(p^{0},r)$ is
formed.\\\indent In the equation (2.2.22) we have obtained the
scattering cross-section for electron type of neutrino, whereas
the scattering cross-section for both muon and tau neutrino is
obtained by using the equation (2.2.17). Now, we proceed to
approximate the expression of scattering cross-section in the
extreme-relativistic as well as non-relativistic limit. In these
two limits the total scattering cross-section in c.g.s unit, for
all three types of neutrino are approximated as
$$\sigma\approx 5.8\times
10^{-50}\times(\frac{E_{ER}}{m_{e}c^{2}})^{2}
\ln(\frac{E_{ER}}{m_{e}c^{2}})\hspace{0.5cm}cm^{2}\hspace{0.5cm}
[extreme-relativistic]\eqno{(2.2.23)}$$
$$\approx 3.44 \times
10^{-49}\times(\frac{E_{NR}}{m_{e}c^{2}})^{\frac{1}{2}}
\hspace{1.7cm}cm^{2}\hspace{0.5cm}[non-relativistic]\eqno{(2.2.24)}$$
It is to be noted that $E_{ER}$ and $E_{NR}$ represent the energy
of the single electron related to the extreme-relativistic and
non-relativistic limits, respectively.\\\indent Now, we check the
goodness of our approximated analytical method. For that purpose
let us obtain the scattering cross-section for electron type of
neutrino in the relativistic case from the equation (2.2.22). It
gives
$$\sigma_{\nu_{e}}\approx 4.16 \times
10^{-51}\times(\frac{E}{m_{e}c^{2}})^{2}
\ln(\frac{E}{2m_{e}c^{2}})\hspace{0.5cm}cm^{2}\eqno{(2.2.25)}$$
where, $E$ is the CM energy.\\ In the Table-2 this result is
compared with the scattering cross-section for electron type of
neutrino obtained by using CalcHep software (version-2.3.7).
\subsection{Calculation of energy loss rate}
A number of different stellar regions are to be taken into account
to calculate the energy loss rate for the electron-neutrino
bremsstrahlung process. We have already calculated the scattering
cross-section for this process and obtained its approximate
expression in the extreme-relativistic and non-relativistic limit.
The later stage of the stellar evolution may be degenerate as well
as non-degenerate depending on the chemical potential. In the
evolution of some stars the electron gas exerts degenerate
pressure which prevents the star from contraction by its
gravitational force. The complete degenerate gas is such in which
all the lower states below the Fermi energy become occupied. There
may exist some ranges of temperature and density where electron
energy is not bounded by Fermi-energy. This non-degeneracy may be
evident for both relativistic and non-relativistic limit, i.e.,
for $\kappa T<m_{e}c^{2}$ and $\kappa T>m_{e}c^{2}$ respectively,
where $\kappa$ is Boltzmann's constant. To calculate the energy
loss rate we use the formula given by
$$\rho\mathcal{E_{\nu}}=\frac{4}{(2\pi)^{6}\hbar^{6}}
\int_{0}^{\infty}\int_{0}^{\infty}\frac{d^{3}p_{1}}{[e^{\frac{E_{1}}{\kappa
T}-\psi}+1]}\frac{d^{3}p_{2}}{[e^{\frac{E_{2}}{\kappa
T}-\psi}+1]}(E_{1}+E_{2})$$$$\times\hspace{0.5cm}\int
d\sigma(E'_{1},E'_{2})g(E'_{1},E'_{2})|\overrightarrow{v_{1}}-\overrightarrow{v_{2}}|
\eqno{(2.3.1)}$$ where $\rho$ is the mass density of the electron
gas and $g(E'_{1},E'_{2})$ stands for Pauli's blocking factor,
given by
$$g(E'_{1},E'_{2})=[1-\frac{1}{e^{\frac{E'_{1}}{\kappa T}-\psi}+1}]
[1-\frac{1}{e^{\frac{E'_{2}}{\kappa T}-\psi}+1}]\eqno{(2.3.2a)}$$
$\psi=\frac{\mu}{\kappa T}$ ($\mu$ represents the chemical
potential of the electron gas) is related to the number density of
the electron by the following formula.
$$n=\frac{2(\kappa T)^{3}}{\pi^{2}(c\hbar)^{3}}
\int_{0}^{\infty}\frac{x[x^{2}-(\frac{m_{e}c^{2}}{\kappa
T})^{2}]^{\frac{1}{2}}} {[e^{x-\psi}+1]}dx\eqno{(2.3.2b)}$$ It is
to be noted that in the equations (2.3.1) and $(2.3.2b)$ the upper
limit of the electron momentum has been taken to infinity; this
may give an impression that the CM energy of the electron is very
high as if it can be comparable to $M_{Z}$ or $M_{W}$. Clearly it
is not true since, practically, the temperature and density of the
stellar core in the later stages of the stellar evolution do not
allow the electron to gain that much energy. Therefore, in the
equations (2.3.1) and $(2.3.2b)$ the upper limit depends on the
temperature and density of the electron gas so that the electron
energy cannot go beyond a certain limit. Now we go through the
following cases.\vspace{0.2cm}\\Case-I:\hspace{0.5cm} In the
extreme-relativistic non-degenerate case the chemical potential
becomes very small compared to $E_{ER}$. In this case the energy
loss rate is calculated as
$$\mathcal{E_{\nu}}\approx5.04\times10^{12}\times T_{10}^{6}
[1+0.82\ln(1.7\hspace{0.1cm}T_{10})]\hspace{0.8cm}erg/gm-sec
\eqno{(2.3.3)}$$ where, $$T_{10}=T\times 10^{-10}$$ From this
analytical expression it is found that the energy loss rate may be
significantly high when the core temperature is more than
$10^{10}$ K. \\Case-II:\hspace{0.5cm} In the extreme-relativistic
degenerate region the density is very high. In case of the neutron
star it reaches to $10^{15}$ $gm/cc$. Pauli's blocking factor
plays an important role to calculate the energy loss rate for
degenerate electron. In extreme-relativistic limit it can be
approximated as
$$\int d\sigma(E'_{1},E'_{2})g(E'_{1},E'_{2})\approx e^{2(1-x_{F})}\sigma\eqno{(2.3.4a)}$$
where $x_{F}$ represents the ratio of the Fermi temperature to the
maximum temperature of the degenerate electron gas at the maximum
density ($\sim 10^{15}$ $gm/cc$). The energy loss rate in the
extreme degenerate case is obtained as
$$\mathcal{E_{\nu}}\approx6.56\times10^{10}\times T_{10}^{6}
[1+0.56\ln(1.7\hspace{0.1cm}T_{10})]\hspace{0.8cm}erg/gm-sec
\eqno{(2.3.4b)}$$ It is high in some degenerate stellar
objects.\\Case-III:\hspace{0.5cm} The non-relativistic effect
becomes important when the central temperature of the star remains
below the $5.93\times 10^{9}$ K and the electron becomes
non-degenerate if $(\frac{\rho}{2})^{\frac{2}{3}}\leq
(\frac{T}{2.97\times 10^{5}K})$. In this case $\mu<m_{e}c^{2}$,
but $\psi$ cannot be neglected as is done in the
extreme-relativistic case. We calculate the energy loss rate as
follows:
$$\mathcal{E_{\nu}}\approx0.88\times10^{-3}\times T_{8}\rho\hspace{0.8cm}
erg/gm-sec\eqno{(2.3.5)}$$ where $T_{8}$ is defined in the same
manner as $T_{10}$. This energy loss rate is not very low in the
region having the temperature $10^{8}-10^{9}$ K and density less
than $10^{6}$ gm/cc, which signifies the importance of this
process in the non-relativistic non-degenerate region.\\\indent It
is worth noting that the energy loss rate in the non-relativistic
degenerate region is very small. Hence this case has not been
considered here.
\subsection{Discussion} Here, we have calculated the scattering
cross-section and obtained its expression both in the relativistic
and non-relativistic limit. In our calculation we have used an
approximation given by the equation (2.2.16). The error arising
out of this approximation is very small. The Table-2 shows that
our result (scattering cross-section for the electron type of
neutrino) is close to that generated by the software. It strongly
supports the approximation that we have used. Cazzola and Saggion
\cite{Cazzola} did not obtain any explicit expression of the
scattering cross-section. Therefore, it is significant to obtain
the analytical expression of the scattering cross-section for this
process. We have calculated the energy loss rate in different
regions characterized by the temperature and density. Our work
shows that the electron-neutrino bremsstrahlung process causes a
large amount of energy loss in the stellar core when the core
temperature $\geq 10^{10}$ K, both in non-degenerate as well as in
degenerate region. In that temperature range the radiation
pressure is so dominating that the gas pressure has negligible
effect \cite{Chandrasekhar}. In this extreme-relativistic region
the process contributes significantly when the electron gas is
non-degenerate. That was clearly shown by Cazzola and Saggion
\cite{Cazzola} by the numerical calculations. But they did not
calculate the energy loss rate in the degenerate region, though
they indicated that the electron-neutrino bremsstrahlung process
might be highly significant in that region. We have also obtained
the expression of the energy loss rate when the electrons are
strongly degenerate. The neutron star, born as a result of type-II
Supernova, is a typical example of the extreme-relativistic
degenerate stellar object. Our study reveals that the energy loss
rate in the non-degenerate region is higher than that in the
degenerate region. This clearly indicates that though during the
neutron star cooling electron-neutrino bremsstrahlung may play a
significant role, the process becomes more important to carry away
the energy from the core of pre-Supernova star, which is a
relativistic non-degenerate stellar object.\\\indent
Non-relativistically, the process becomes significant when the
temperature attains $10^{8}$ K. At this temperature the burning of
helium gas in the stellar core takes place \cite{Hayashi}. In the
temperature range $10^{8}- 10^{9}$ K the gas pressure is
dominating over the radiation pressure, though the effect of
radiation pressure cannot be neglected. In addition to that the
region will be non-degenerate if the density $<2\times 10^{6}$
$gm/cc$. The electron-neutrino bremsstrahlung process may have
some effect in this region though the energy loss rate is not so
high as it is in the extreme-relativistic case. Thus we find the
process is important in non-degenerate region, especially when the
electrons are highly relativistic. The process may also have
significant effect for the degenerate electron gas only when the
temperature and density of the electron gas is high enough.
Therefore, we can say that the electron-neutrino bremsstrahlung is
an important mechanism for the energy loss in the stellar core
during the late stages of stellar evolution.
\subsection{Appendix}
The trace rule for the $\gamma$ matrices is given as follows:
$$\frac{1}{4}Tr[(a_{1}^{\tau}\gamma_{\tau})(a_{2}^{\tau}\gamma_{\tau})
(a_{3}^{\tau}\gamma_{\tau}).........(a_{n}^{\tau}\gamma_{\tau})]=
(a_{1}a_{2})Tr[(a_{3}^{\tau}\gamma_{\tau})(a_{4}^{\tau}\gamma_{\tau})
.........(a_{n}^{\tau}\gamma_{\tau})]$$
$$-(a_{1}a_{3})Tr[(a_{2}^{\tau}\gamma_{\tau})(a_{3}^{\tau}\gamma_{\tau})
.........(a_{n}^{\tau}\gamma_{\tau})]+ .........
+(a_{1}a_{n})Tr[(a_{2}^{\tau}\gamma_{\tau})(a_{3}^{\tau}\gamma_{\tau})
.........(a_{n-1}^{\tau}\gamma_{\tau})]\eqno{(1)}$$ We construct
the following formula with the aid of this trace rule.
$$\frac{1}{4}Tr[(p_{1}^{\tau}\gamma_{\tau}+m_{e})\gamma_{\alpha}(P^{\tau}\gamma_{\tau}+m_{e})
\gamma_{\rho}(p_{1}'^{\tau}\gamma_{\tau}+m_{e})\gamma_{\sigma}(Q^{\tau}
\gamma_{\tau}+m_{e})\gamma_{\beta}]\vspace{0.2cm}=$$
$$[g_{\alpha\beta}g_{\rho\sigma}-g_{\alpha\sigma}g_{\rho\beta}+
g_{\alpha\rho}g_{\sigma\beta}][(p_{1}P)(p_{1}'Q)+(p_{1}Q)(p_{1}'P)-
(p_{1}p_{1}')(PQ)$$
$$+m_{e}^{2}\{(p_{1}p_{1}')+(PQ)-(p_{1}P)-(p_{1}'P)-(p_{1}Q)-(p_{1}'Q)\}+m_{e}^{4}]$$
$$+$$
$$[(p_{1}'P)-m_{e}^{2}]
[g_{\sigma\beta}(p_{1\alpha}Q_{\rho}-p_{1\rho}Q_{\alpha})+
g_{\rho\beta}(p_{1\sigma}Q_{\alpha}-p_{1\alpha}Q_{\sigma})+
g_{\alpha\sigma}(p_{1\beta}Q_{\rho}+p_{1\rho}Q_{\beta})$$
$$-g_{\alpha\rho}(p_{1\sigma}Q_{\beta}+p_{1\beta}Q_{\sigma})-g_{\rho\sigma}
(p_{1\alpha}Q_{\beta}+p_{1\beta}Q_{\alpha})+
g_{\alpha\beta}(p_{1\rho}Q_{\sigma}-p_{1\sigma}Q_{\rho})]$$
$$+$$
$$[(p_{1}'Q)-m_{e}^{2}]
[-g_{\sigma\beta}(p_{1\alpha}P_{\rho}+p_{1\rho}P_{\alpha})+
g_{\rho\beta}(p_{1\sigma}P_{\alpha}+p_{1\alpha}P_{\sigma})-
g_{\alpha\sigma}(p_{1\beta}P_{\rho}-p_{1\rho}P_{\beta})$$
$$-g_{\alpha\rho}(p_{1\sigma}P_{\beta}-p_{1\beta}P_{\sigma})-
g_{\rho\sigma}(p_{1\alpha}P_{\beta}+p_{1\beta}P_{\alpha})-
g_{\alpha\beta}(p_{1\rho}P_{\sigma}-p_{1\sigma}P_{\rho})]$$
$$+$$
$$[(p_{1}P)-m_{e}^{2}]
[-g_{\sigma\beta}(p_{1\alpha}'Q_{\rho}-p_{1\rho}'Q_{\alpha})+
g_{\rho\beta}(p_{1\sigma}'Q_{\alpha}+p_{1\alpha}'Q_{\sigma})+
g_{\alpha\sigma}(p_{1\beta}'Q_{\rho}-p_{1\rho}'Q_{\beta})$$
$$-g_{\alpha\rho}(p_{1\sigma}'Q_{\beta}+p_{1\beta}'Q_{\sigma})+
g_{\rho\sigma}(p_{1\alpha}'Q_{\beta}-p_{1\beta}'Q_{\alpha})-
g_{\alpha\beta}(p_{1\rho}'Q_{\sigma}+p_{1\sigma}'Q_{\rho})]$$
$$+$$
$$[(p_{1}Q)-m_{e}^{2}]
[-g_{\sigma\beta}(p_{1\alpha}'P_{\rho}+p_{1\rho}'P_{\alpha})-
g_{\rho\beta}(p_{1\sigma}'P_{\alpha}-p_{1\alpha}'P_{\sigma})+
g_{\alpha\sigma}(p_{1\beta}'P_{\rho}+p_{1\rho}'P_{\beta})$$
$$+g_{\alpha\rho}(p_{1\sigma}'P_{\beta}-p_{1\beta}'P_{\sigma})-
g_{\rho\sigma}(p_{1\alpha}'P_{\beta}-p_{1\beta}'P_{\alpha})-
g_{\alpha\beta}(p_{1\rho}'P_{\sigma}+p_{1\sigma}'P_{\rho})]$$
$$+$$
$$[(p_{1}p'_{1})-m_{e}^{2}]
[g_{\sigma\beta}(P_{\alpha}Q_{\rho}+P_{\rho}Q_{\alpha})-
g_{\rho\beta}(P_{\sigma}Q_{\alpha}+P_{\alpha}Q_{\sigma})-
g_{\alpha\sigma}(P_{\beta}Q_{\rho}+P_{\rho}Q_{\beta})$$
$$+g_{\alpha\rho}(P_{\sigma}Q_{\beta}+P_{\beta}Q_{\sigma})-
g_{\rho\sigma}(P_{\alpha}Q_{\beta}-P_{\beta}Q_{\alpha})-
g_{\alpha\beta}(P_{\rho}Q_{\sigma}-P_{\sigma}Q_{\rho})]$$
$$+$$
$$[(PQ)-m_{e}^{2}]
[-g_{\sigma\beta}(p_{1\alpha}'p_{1\rho}-p_{1\rho}'p_{1\alpha})-
g_{\rho\beta}(p_{1\sigma}'p_{1\alpha}+p_{1\alpha}'p_{1\sigma})-
g_{\alpha\sigma}(p_{1\beta}'p_{1\rho}+p_{1\rho}'p_{1\beta})$$
$$+g_{\alpha\rho}(p_{1\beta}'p_{1\sigma}-p_{1\sigma}'p_{1\beta})+
g_{\rho\sigma}(p_{1\alpha}'p_{1\beta}+p_{1\beta}'p_{1\alpha})+
g_{\alpha\beta}(p_{1\rho}'p_{1\sigma}+p_{1\sigma}'p_{1\rho})]$$
$$+$$
$$[(p_{1\sigma}'p_{1\alpha}+p_{1\alpha}'p_{1\sigma})
(P_{\beta}Q_{\rho}+P_{\rho}Q_{\beta})+
(p_{1\beta}'p_{1\rho}+p_{1\rho}'p_{1\beta})
(P_{\sigma}Q_{\alpha}+P_{\alpha}Q_{\sigma})+$$
$$(p_{1\rho}'p_{1\alpha}-p_{1\alpha}'p_{1\rho})
(P_{\sigma}Q_{\beta}+P_{\beta}Q_{\sigma})+
(p_{1\sigma}'p_{1\beta}-p_{1\beta}'p_{1\sigma})
(P_{\alpha}Q_{\rho}+P_{rho}Q_{\alpha})+$$
$$(p_{1\alpha}'p_{1\beta}+p_{1\beta}'p_{1\alpha})
(P_{\rho}Q_{\sigma}-P_{\sigma}Q_{\rho})+
(p_{1\rho}'p_{1\sigma}+p_{1\sigma}'p_{1\rho})
(P_{\alpha}Q_{\beta}-P_{\beta}Q_{\alpha})]\eqno{(2)}$$ where,
$$P=q+p_{1}'$$  $$Q=p_{1}-q$$
In order to evaluate the various trace terms the formula
constructed in the equation (2) is very much helpful.
\section{Neutrino Synchrotron Radiation}
\subsection{Introduction}
In the stellar interior the neutrino emission can be greatly
enhanced by the presence of a high magnetic field. When the core
of the star contracts considerably the strength of the magnetic
field existing in the stellar interior is increased. The fast
moving electron not only emits photon as in the case of
synchrotron radiation, but it may also emit neutrinos, when the
electron gets spiraled along the magnetic field lines. The
influence of a high magnetic field on the free electron causes the
emission of neutrino-antineutrino pair and the process is called
neutrino synchrotron radiation (NSR) in analogy with the
electromagnetic synchrotron radiation (ESR) in which
electromagnetic radiation occurs by a magnetically accelerated
electron. Emission of the ESR or ordinary synchrotron radiation
was first discovered as a by-product in the motion of electrons in
circular accelerator synchrotron and hence the name of the
radiation. It was found that, at relativistic speed of the
electrons in the magnetic field of the accelerator, synchrotron
radiation plays a fundamental role; precisely, it determines the
dynamics of the particles in the accelerator: radiative energy
losses, the classical radiative damping of betatron oscillations
and the quantum fluctuations of the particle trajectories. White
dwarves are the end stages of low mass stars. In some white
dwarves having the core density $1.6\times 10^{7} - 10^{10}$
$gm/cc$ and temperature $10^{7} - 5.5\times 10^{7}$ K the magnetic
field with intensity $10^{9} - 10^{11}$ G may exist. This value is
high in such kind of degenerate stellar object. A strong magnetic
field, which is very much relevant in the context of our
discussion, may be present also in the neutron stars. The neutron
star is a dense stellar structure formed as the result of type-II
supernova and in the newly born neutron star it would have central
density near about $10^{15}$ $gm/cc$ and temperature more than
$10^{11}$ K. In some of the neutron stars the high magnetic field
($10^{12} - 10^{16}$ G) may be found. It is found in a few cases
that the highly relativistic degenerate stellar objects like
neutron stars are magnetized.\\\indent In 1967 Landstreet
\cite{Landstreet}, first time, considered the neutrino synchrotron
radiation process under the hypothesis that there would exist a
direct $e-\nu_{e}$ coupling. He computed, approximately, the
neutrino radiation from a completely relativistic electron gas in
presence of a large magnetic field. He pointed that the process
would have astrophysical significance in the evolution of stars
with large electron energy and potentially large magnetic field,
such as white dwarves; but the computation of the total neutrino
luminosity of several model of white dwarves showed that the
neutrino luminosity would be limited by electron degeneracy to
values much less than photon luminosity. It is found from
Landstreet's calculation that for white dwarf with magnetic field
$H\leq 10^{11}$ G, the NSR luminosity is smaller than the photon
luminosity by a factor of $10^{3}$ or more, whereas that is
smaller than the photon luminosity by a factor of $10^{7}$ or more
when $H\leq 10^{9}$ G. It is evident from his work that the
process might have some effect in the neutron star where the
magnetic field would be very high and the stellar core is highly
relativistic as well as degenerate. In 1970 Canuto et al.
\cite{Canuto1} reviewed the neutrino synchrotron process to
calculate the neutrino luminosity due to completely relativistic
gas in the framework of V-A theory of the universal weak Fermi
interaction, by standard field theoretic method. In high density
their result agreed to that of Landstreet \cite{Landstreet}.
Raychaudhuri \cite{Raychaudhuri3} considered this process
according to the photon neutrino weak coupling theory in presence
of a strong magnetic field for completely relativistic electron
gas. He showed \cite{Raychaudhuri3} that in the framework of
photon neutrino weak coupling theory for any white dwarf with
$T\leq 5. 5\times 10^{7}$ K and $H\leq 10^{11}$ G, the NSR
luminosity is greater than the photon luminosity by a factor
$10^{3}$ or more; even for $H\leq 10^{9}$ G the NSR luminosity is
of higher order than the photon luminosity. According to the
photon neutrino weak coupling theory, this process might be
significant during the cooling of white dwarf. In his subsequent
paper \cite{Raychaudhuri3} he indicated that by the application of
this theory the NSR process alone should be responsible for
cooling of the core of neutron star. After that, a number of works
\cite{Vidaurre, Kaminker1991, Kaminker1992, Yakovlev, Bezchastnov}
were carried out on the NSR process relating to the stellar
physics. Out of these the paper of Bezchastnov et al.
\cite{Bezchastnov} draws our particular attention. They studied
the NSR process by relativistic degenerate electrons in presence
of a strong magnetic field with the emphasis on the electron
transitions associated with one or several cyclotron harmonics.
They calculated the luminosity using quantum as well as
quasi-classical treatment and showed the NSR emissivity might give
dominant contribution at a strong magnetic field in the high
density layers of the neutron star crusts. Their result indicated
that the NSR would play a significant role in the cooling theories
of magnetized neutron stars. All these works, therefore, clearly
indicate that the NSR could be thought as an important process for
neutrino energy loss in the later stages of the stellar evolution
when the stellar objects are highly magnetized.\\\indent We are
strongly motivated to calculate this process according to the
Standard Model. In the framework of electro-weak theory each
neutrino-antineutrino pair is emitted via exchange of the
intermediate $Z$-boson. Though in NSR the scattering process
occurs between a particle and a field, still it can be considered
and calculated in the framework of electro-weak interaction
theory. It is significant to calculate the scattering
cross-section for this process which may reflect some information
regarding it. It is also worth noting that the process is
considered in the highly relativistic and degenerate electron gas,
when a strong magnetic field is present. We also calculate the
energy loss rate and then the NSR luminosity in the white dwarf
and neutron star. Our result is compared with Landstreet's result
\cite{Landstreet} and its astrophysical significance is discussed
briefly. \footnote {This result is published in {\it Astroparticle
physics} \cite{Bhattacharyya3}.}
\subsection{Calculation of scattering cross-section}
It has already been mentioned that in the neutrino synchrotron
process an electron is scattered by a magnetic field to release
the neutrino-antineutrino pair, and thus the NSR radiation takes
place. Let us consider a situation where an electron travels in
high momentum, or in other words, it has highly relativistic
energy, in a strong magnetic field. This phenomenon is represented
by a special choice of minimal coupling. The magnetic field
compels the electron to move in a spiral orbit. The energy of the
electron is quantized in the direction perpendicular to the field
$H$. Without any loss of generality the direction of magnetic
field $H$ is taken along the $z$-axis. Due to that quantization of
energy the components of transverse momentum $p_{x}$ and $p_{y}$
are replaced by the field strength $H$, while the longitudinal
component $p_{z}$ remains unaffected. If we consider $p_{1}^{0}$
and $p_{2}^{0}$ be the energy of the electron just before and
after the release of neutrino-antineutrino pair, respectively, in
presence of a magnetic field, we can write
$$(p_{1}^{0})^{2}=(p_{\parallel}^{0})^{2}+2n_{1}\frac{H}{H_{c}}m_{e}^{2}\eqno {(3.2.1)}$$
$$(p_{2}^{0})^{2}=(p_{\parallel}^{0})^{2}+2n_{2}\frac{H}{H_{c}}m_{e}^{2}\eqno {(3.2.2)}$$
where,
$$(p_{\parallel}^{0})^{2}=m_{e}^{2}+p_{z}^{2}\eqno {(3.2.3)}$$
To be noted that $n_{1}$ and $n_{2}$ are the Landau levels for the
incoming and outgoing electrons, respectively. This Landau level
is very much analogous to the principal quantum number in the
Bohr's atomic theory. Here the terms `incoming' and `outgoing' are
used in the sense of `before' and `after' release of the neutrino
pair. The range of $n_{2}$ should be $0, 1, 2 ...n_{1}$. It is
quite obvious that $n_{1}$ does not depend on $p^{0}_{\parallel}$,
but depends on the strength of the magnetic field. Here $H_{c}$ is
called critical magnetic field having the value $4.414\times
10^{13}$ G. In the quasi-classical limit the Landau level $n$ is
related to the transverse component of the momentum by the
relation
$$(\frac{2n}{eH})^{\frac{1}{2}}=\frac{p_{\perp}}{eH}=R_{n}\eqno{(3.2.4)}$$
where $R_{n}$ is called Larmor radius. Bezchastnov {\it et al.}
\cite{Bezchastnov} discussed this quasi-classical treatment in the
neutrino synchrotron process and carried out the neutrino
emissivity numerically. It is convenient to start our study with
ordinary QED process and we, at first, intend to construct the
matrix element for the ordinary synchrotron radiation given by
$$e^{-}\longrightarrow^{magnetic field} e^{-}+\gamma$$ Let us
consider the case when an electron is scattered by an
electromagnetic field characterized by the electromagnetic vector
potential $A_{\rho}(k)$. The scattering amplitude can be
constructed \cite{Greiner1} according to the Feynman rule as
follows:
$$M_{fi}^{em}=-ie[(\overline{u}(p_{2})\gamma^{\rho}A_{\rho}(k)u(p_{1})]\eqno{(3.2.5)}$$
In particular, $A_{0}=-\frac{4\pi Ze}{|\overrightarrow{q}|^{2}}$
 and $\overrightarrow{A}=0$ represent well known
coulomb scattering. It is to be noted that here the scattering
process takes place in presence of an external field. In the same
manner we try to construct the matrix element for the synchrotron
process. In the matrix element for electromagnetic synchrotron
radiation process the term $A_{\rho}(k)$ will be replaced by the
polarization four vector $\varepsilon_{\rho}$ indicating the
outgoing photon, since the photon is emitted by the ESR. Thus the
matrix element for the ESR can be constructed as
$$M_{fi}'=-ie(\frac{H}{H_{c}})^{\frac{1}{2}}[(\overline{u}(p_{2})\gamma^{\rho}u(p_{1})]
\varepsilon_{\rho}(q)\eqno{(3.2.6)}$$ Here the factor
$(\frac{H}{H_{c}})^{\frac{1}{2}}$ indicates that in absence of a
magnetic field no transition of electron is possible. This factor
has some similarity with dimensionless coupling constant
$\sqrt{\alpha}$ in QED, though $(\frac{H}{H_{c}})^{\frac{1}{2}}$
is not a constant at all, but depends on the field strength $H$.
It is clear that the transition of electron from one Landau level
to the next lower one is due to the emission of photon.\\\indent
In this section we concentrate on the neutrino synchrotron process
in which the neutrino-antineutrino pair is emitted instead of the
emission of photon. We consider that the electron moving in a
magnetic field has highly relativistic longitudinal component of
the momentum and this component is not affected throughout the
process. In other words, the longitudinal component $p_{z}$
remains same before and after the emission of neutrino pair.
Figure-8 represents the Feynman diagram for the NSP, while the
scattering amplitude of this process may take the form
$$ M_{fi}=\frac{ig^{2}}{(4\cos\theta_{W})^{2}
M_{Z}^{2}}(\frac{H}{H_{c}})^{\frac{1}{2}}[\overline{u}(p_{2})\gamma^{\rho}
(g'_{V}-\gamma_{5})u(p_{1})]
[v_{\nu}(q_{2})\gamma_{\rho}(1-\gamma_{5})u_{\nu}(q_{1})]\eqno{(3.2.7)}$$
where,
$$g'_{V}=1-4\sin^{2}\theta_{W}$$
Let us calculate the sum of $|M_{fi}|^{2}$ over the spin as
follows.
$$\frac{1}{2}\sum|M_{fi}|^{2}=\frac{g^{4}}{M_{Z}^{4}m_{\nu}^{2}(4\cos\theta_{W})^{4}}
(\frac{H}{H_{c}})X^{\rho\sigma}(p_{1},p_{2})Y_{\rho\sigma}(q_{1},q_{2})\eqno{(3.2.8)}$$
where,
$$X^{\rho\sigma}(p_{1},p_{2})=p_{1}^{\rho}p_{2}^{\sigma}+p_{1}^{\sigma}p_{2}^{\rho}
+\{m_{e}^{2}-(p_{1}p_{2})g^{\rho\sigma}\}\eqno{(3.2.9)}$$ and
$$Y_{\rho\sigma}(q_{1},q_{2})=2[q_{1\rho}q_{2\sigma}+q_{1\sigma}q_{2\rho}
-(q_{1}q_{2})g_{\rho\sigma}+iq_{1}^{\alpha}q_{2}
^{\beta}\varepsilon_{\alpha\beta\rho\sigma}]\eqno{(3.2.10)}$$ We
have already stated that our aim is to calculate the scattering
cross-section for the NSR process. Basically, the scattering
cross-section can be defined as the transition rate per unit of
incident flux; the likelihood of any particular final state can be
expressed in terms of such cross-section. For NSR process the
scattering cross-scetion is related to the matrix element by the
following relation.
$$\sigma=\frac{\mathcal{S}}{\mid \overrightarrow{v}\mid}\frac{2m_{e}}{2p_{1}^{0}}
\sum_{n_{2}=0}^{n_{1}}\frac{2m_{e}}{2p_{2}^{0}}\int\frac{2m_{\nu}}{2q_{1}^{0}}
\frac{d^{3}q_{1}}{(2\pi)^{3}}\frac{2m_{\nu}}{2q_{2}^{0}}
\frac{d^{3}q_{2}}{(2\pi)^{3}}(2\pi)^{4}\delta^{4}(p_{1}-p_{2}-q_{1}-q_{2})$$$$
\hspace{7cm}\frac{1}{2}\sum \mid
M_{fi}(p_{1},p_{2},q_{1},q_{2})\mid^{2}\eqno{(3.2.11)}$$ where,
$\mathcal{S}$ is the degeneracy factor. \\\indent Let us now
reconsider the physics behind the neutrino synchrotron process.
The electron moving in a spiral path changes the Landau level one
by one after emitting neutrino-antineutrino pair at every level.
Instead of considering each individual level we consider initial
and final level. Each individual Landau level does not contain any
information in the scattering theory that we have developed. So we
put
$$n_{1}=n \hspace{2cm}[p_{1}^{0}=p^{0}]$$ and
$$n_{2}=0 \hspace{2cm}[p_{2}^{0}=p_{\parallel}^{0}]$$.
\indent In the scattering process, though there is no exact
localized scattering centre, we consider the point, in
approximation, where the electron jumps from the $n$th state to
the ground state to release the neutrino-antineutrino pair, as the
scattering centre. Considering all three types of neutrino the
expression for scattering cross-section takes the form as follows.
$$\sigma=\frac{3g^{4}}{32(2\pi)^{2}M_{Z}^{4}}(\frac{H}{H_{c}})\frac{X^{\rho\sigma}
(p_{1},p_{2})I_{\rho\sigma}(q)}{p_{1}^{0}p_{2}^{0}|\overrightarrow{v}|}\eqno{(3.2.12)}$$
where,
$$I_{\rho\sigma}(q)=\int\int\delta^{4}(q-q_{1}-q_{2})\frac{Y_{\rho\sigma}
(q_{1},q_{2})}{q_{1}^{0}q_{2}^{0}}d^{3}q_{1}d^{3}q_{2}\eqno{(3.2.13)}$$
and
$$q=p_{1}-p_{2}=q_{1}+q_{2}\eqno{(3.2.14)}$$
We are now going to evaluate the integral $I_{\rho\sigma}(q)$
given by the equation (3.2.13). This integral depending on the
four vector $q$ is a rank-2 tensor, and therefore, its most
general form becomes
$$ I_{\rho\sigma}(q)=Aq^{2}g_{\rho\sigma}+Bq_{\rho}q_{\sigma}\eqno{(3.2.15)}$$
We can calculate the dimensionless quantity $A$ and $B$ in the
frame where $\overrightarrow{q}=0$. Here an important point is to
be noted that the mass of the neutrino can be ignored with respect
to its energy, i.e.,
$$q^{0}=q^{0}_{1}+q^{0}_{2}\gg m_{\nu}$$
Taking it into account we can obtain
$$-A=B=\frac{2\pi}{3}\eqno{(3.2.16)}$$
and putting these into the equation (3.2.15) we get the complete
expression for $I_{\rho\sigma}(q)$. We can also find the
expression for $X^{\rho\sigma} (p_{1},p_{2})$ from the equation
(3.2.9) and so it is now easy to obtain the scalar product
$X^{\rho\sigma}I_{\rho\sigma}$ given in the equation (3.2.12).
Obtaining this term and simplifying the expression we evaluate the
scattering cross-section. We like to obtain the scattering
cross-section in C.G.S. unit and for that purpose we introduce the
conversion factor $(c\hbar)^{2}$. Thus the scattering
cross-section is obtained as
$$ \sigma\approx
2.06\times10^{-46}(\frac{H}{H_{c}})\frac{1}{|\overrightarrow{v}|}(\frac{p_{\perp}c}
{E_{\parallel}})^{2}f_{n}(p_{\perp}c,E_{\parallel})\hspace{0.5cm}cm^{2}\eqno{(3.2.17)}$$
where,
$$f_{n}(p_{\perp}c,E_{\parallel})=[1+\frac{1}{4}(\frac{p_{\perp}c}
{E_{\parallel}})^{2}+2(\frac{m_{e}c^{2}}{E_{\parallel}})^{2}+....]
[1-(\frac{m_{e}c^{2}}{E_{\parallel}})^{2}]^{\frac{-1}{2}}
[1+(\frac{p_{\perp}c}{E_{\parallel}})^{2}]^{\frac{-1}{2}}\eqno{(3.2.18)}$$
$E_{\parallel}$ and $p_{\perp}$ are defined as follows:
$$E_{\parallel}=p_{\parallel}^{0}c\eqno{(3.2.19)}$$ and
$$p_{\perp}^{2}=2n\frac{H}{H_{c}}m_{e}^{2}c^{2}\eqno{(3.2.20)}$$
which is taken to be completely independent parameter in our
development. The theory will be consistent if $p_{\perp}c<
E_{\parallel}$, i.e., the maximum value of the energy generated
due to the magnetic field does not exceed the amount
$E_{\parallel}$ given by
$$E_{\parallel}^{2}=m_{e}^{2}c^{4}+p_{z}^{2}c^{2}\eqno{(3.2.21)}$$
To be noted that we obtain the scattering cross-section given by
the equation (3.2.17) assuming the electron is ultra-relativistic.
In the neutrino synchrotron process this consideration does not
violate the generality of the theory. For simplification it is
taken that
$$f_{n}(p_{\perp}c,E_{\parallel})\simeq1\eqno{(3.2.22)}$$ and thus we
can simplify the expression of the scattering cross-section.
\subsection{Calculation of energy loss rate and luminosity}
We are going to study the NSR process to verify whether it is
important during the evolution of star, especially in the later
stages. Therefore, we should calculate how much energy can be
radiated per unit mass per second by this process. In calculating
the energy loss rate it is important to consider $p_{\perp}$ and
$p_{z}$ as two independent parameters. The scattering
cross-section contains both of these terms. We are to take the
summation over the Landau level $n$, but here we replace discrete
summation by continuous integration over $p_{\perp}$
\cite{Bezchastnov}. The energy loss rate can be obtained from the
following relation:
$$\rho\mathcal{E}_{\nu}=\int
d\overrightarrow{N}(p_{\perp})d\overrightarrow{N}(p_{z})E\sigma|\overrightarrow{v}|
\eqno{(3.3.1)}$$ where $d\overrightarrow{N}(p_{\perp})$ and
$d\overrightarrow{N}(p_{z})$ are the momentum distribution given
by
$$d\overrightarrow{N}(p_{\perp})=\frac{2}{\pi^{2}\hbar^{3}}p_{\perp}^{2}
dp_{\perp}\eqno{(3.3.2)}$$
$$d\overrightarrow{N}(p_{z})=\frac{2}{\pi^{2}\hbar^{3}}\frac{p_{z}^{2}dp_{z}}
{[e^{\frac{E_{\parallel}-E_{F}}{\kappa T}}+1]}\eqno{(3.3.3)}$$ and
$$E=p^{0}c$$ Here $E$ is the total energy of the electron just before
emission of the neutrino-antineutrino pair. To be noted that
$p^{0}$ is given by the relation
$$(p^{0})^{2}=(p_{\parallel}^{0})^{2}+p_{\perp}^{2}$$
 It is assumed that $p_{\perp}$ related to the Landau level does not exceed
$p_{\parallel}^{0}$. Here we exploit the fact that the electron is
ultra-relativistic, i.e., $p_{z}\gg m_{e}c$; it is also assumed
that the electron is degenerate, i.e., the energy of the electron
due to its linear motion always stays behind the Fermi energy
level $E_{F}$. Finally, the energy loss rate is calculated as
$$\mathcal{E}_{\nu}=2.25\times 10^{-1}(\frac{H}{H_{c}})T_{7}^{6}\rho^{\frac{-2}{3}}
\hspace{0.5cm}erg/gm-sec\eqno{(3.3.4)}$$ where,
$$T_{7}=T \times 10^{-7}$$
From the Landstreet's result \cite{Landstreet} we can find
$$\mathcal{E}^{L}_{\nu}=2.32\times
10^{-3}(\frac{H}{H_{c}})^{\frac{2}{3}}
T_{7}^{\frac{19}{3}}\rho^{\frac{-5}{9}}\hspace{0.5
cm}erg/gm-sec\eqno{(3.3.5)}$$ These two results are plotted and
compared graphically both for the white dwarf and neutron star to
observe the deviations. In the Figure-9 the energy loss rate
(obtained by our method and also from Landstreet's result) is
plotted against the temperature varying from $10^{7}$ K to
$5.5\times 10^{7}$ K at the density $10^{8}$ $gm/cc$,
characterizing the white dwarf. The magnetic field is taken to be
$10^{11}$ G, a maximum possible magnetic field present in the
white dwarf. In this comparison it is found that the two graphs
almost coincide, i.e., our result is very close to that of
Landstreet. The same comparison has been carried out for neutron
star in the Figure-10. In this case the range of temperature is
taken as $6\times 10^{9}- 10^{10}$ K at the maximum possible
density ($10^{15}$ $gm/cc$) and the critical magnetic field
($4.414\times 10^{13}$ G). Here, also, our result is close to that
of Landstreet, although our approach to obtain the expression of
energy loss rate is different from that adopted by Landstreet. The
neutrino luminosity expressed in the unit of solar luminosity is
obtained for our case as well as for Landstreet's calculations. In
Table-3 we have computed the neutrino luminosity by the NSR in a
white dwarf having central density $10^{8}$ $gm/cc$ at the
magnetic field $10^{11}G$, in the temperature range $10^{7}-
5.5\times 10^{7}$ K. The neutrino luminosity obtained in this
range is very low. In Table-4 the neutrino emissivity is
obtained in the neutron star, at the magnetic field $H=H_{c}=4.414
\times 10^{13} $ G, at the density $\rho=10^{15}$ $gm/cc$,
$10^{14}$ $gm/cc$, $10^{13}$ $gm/cc$ and in the temperature range
$6\times 10^{9}- 10^{10}$ K. This table indicates that the NSR
luminosity is significantly high in case of neutron star. The NSR
is considered in the degenerate stellar object. In the
non-degenerate region the neutrino synchrotron process is not
significant.
\subsection{Discussion}
An important aspect of our work is the calculation of scattering
cross-section, which reflects, no doubt, a new approach in
comparison with the earlier works on the NSR process. Since,
earlier, it was not attempted to calculate the process in the
Standard Model, there was indeed no scope for considering the fact
that the NSR would occur through the exchange of the intermediate
boson. We have carried out the calculation in the framework of
electro-weak theory and the matrix element, we have constructed,
contains the expression for intermediate Z-boson. The
$z$-component of the electron momentum $p_{z}$ is not affected by
the influence of the magnetic field and therefore, the neutrino
pair emission takes place only due to the change of Landau levels,
i.e., the change in $p_{\perp}$. It is obvious that both the NSR
and ESR occur simultaneously during stellar evolution, but the
`weak' one might dominate when the electron would be
extreme-relativistic and degenerate in nature.
\\\indent Canuto {\it et al.} \cite{Canuto1} suggested that at the low
density their results would have some discrepancies compared to
that of Landstreet \cite{Landstreet}. This is, perhaps, due to the
different approximations used by different authors. Anyway, this
is beyond the scope of our verification as we are concentrating
solely on the neutrino emission in the later phases of stellar
evolution where the density is high enough. If we study Figure-9
we observe that our result matches with that of Landstreet
\cite{Landstreet}, although Table-3 shows that the NSR gives
very low neutrino luminosity. Comparatively low temperature $(\sim
10^{7}$ K) in the white dwarf core is the reason for such kind of
low neutrino luminosity. Therefore, the process has very little
significance in the white dwarf, even in presence of a high
magnetic field ($\sim 10^{11}$ G). The scenario changes a bit when
the neutrino luminosity is calculated in the neutron star.
Table-4 indicates that in the temperature range $6\times 10^{9}
- 10^{10}$ K the NSR luminosity is very high in the neutron star.
Thus this process may play a significant role in the cooling of
highly magnetized neutron star. Therefore, we can say the neutrino
synchrotron process is important during the later phase of the
stellar evolution, especially in the degenerate,
extreme-relativistic and highly magnetized stellar objects.

\section{Neutrino Bremsstrahlung Process in Magnetic Field}
\subsection{Introduction}
In 1959 it was pointed out by Pontecorvo \cite{Pontecorvo} that
neutrino-antineutrino pair could be emitted when electron collides
with a heavy nucleus. He calculated the rate of the probabilities
for the emission of photon as well as neutrino pair by the
bremsstrahlung radiation and concluded that at certain stages of
stellar evolution it might be possible that the energies sent into
space in the form of photons and neutrinos would be comparable, in
spite of very low rate. He noted further that due to strong
temperature dependence of the probability of the process of
neutrino bremsstrahlung radiation and also with the increase of
the nuclei $Z$, resulting into the decrease of photon mean free
path, the importance of the neutrino bremsstrahlung process would
be greater for energy balance. All these would lead to the
supposition that the neutrino bremsstrahlung process could be
important in certain stages of stellar evolution at which the
temperature and average $Z$ (nuclei) considerably exceed the
corresponding values for the sum. In 1960 Gandel'man and Pinaev
\cite{Gandel'man} investigated this process for non-relativistic
non-degenerate electron gas in the context of the intended
astrophysical application. They showed that in a certain range of
high density and high temperature the energy loss by
bremsstrahlung emission of neutrino pairs would become greater
than the loss due to radiative thermal conductivity. They
calculated the scattering cross-section and then neutrino
luminosity, and the comparison was done with respect to the photon
luminosity. Such comparison indicated that the neutrino luminosity
could produce an appreciable effect only in the stars with high
density. They speculated that the bremsstrahlung process would
lead to even greater energy loss than that by URCA process. Though
they did not investigate the regime of the degenerate equation of
state, but from quantitative considerations they concluded that
the neutrino effect might play an important role in higher density
than that in the non-degenerate regime. In 1969 Festa and Ruderman
\cite{Festa} extended the earlier calculations of neutrino
bremsstrahlung to the relativistic and degenerate region
considering the screening of the coulomb field, important at high
density, and also assuming some lattice effects. In the limit of
infinite density \footnote {That implies the density is much
higher than $10^{6}$ $gm/cc$; in the neutron star it may go to
$10^{15}$ $gm/cc$.} where the electrons are very much
relativistic, their calculations indicated, the energy loss rate
would depend on the temperature. It was then pointed out that the
neutrino bremsstrahlung process must have maximum effect at the
high density and moderate temperature. In 1971 Saha \cite{Saha}
studied this process in the framework of photon neutrino weak
coupling theory and compared his result with that obtained in the
current current coupling theory. He finally concluded that the
process would be important only in the cases of highly dense stars
having the temperature below $10^{7}$ K. Cazzola et al.
\cite{Cazzola2} carried out the detailed calculations of neutrino
bremsstrahlung numerically to cover the full range of electron
degeneracy, and it supported the work of Festa and Ruderman
\cite{Festa}. In 1973 Flowers \cite{Flowers} studied the process
extensively using a many-body approach and got the similar
results. In 1976 Dicus et al. \cite{Dicus} reconsidered the
process for the relativistic degenerate electron gas in the
framework of electro-weak theory including the neutral current.
They carried out the calculations related to this process
considering weak as well as strong screening effect, which would
not change the energy loss rate by more than a factor of 2. Their
result revealed that in the Standard Model the energy loss rate
would become 1.3 times the rate in V-A theory.\\\indent The works
of most of the researchers indicate that the process has got
maximum effect when the electron gas is highly degenerate and
relativistic. Even though Gandel'man and Pinaev \cite{Gandel'man}
calculated the process in the non-degenerate non-relativistic
region, still they comprehended the process would become important
in the degenerate case. The calculation of this process undergone
by Festa and Ruderman \cite{Festa}, and then by Dicus et al.
\cite{Dicus} is important in the relativistic degenerate regime
when the density is very high. It is known that the region in
which the neutrino bremsstrahlung process is supposed to have the
maximum effect, may be highly magnetized in a few cases. We are
going to study the neutrino bremsstrahlung process in presence of
a magnetic field. We are strongly motivated by the supposition
that the bremsstrahlung process may be influenced by the high
magnetic field, present in some neutron stars and magnetars. In
the previous chapter we have discussed that a strong magnetic
field causes neutrino-antineutrino emission through neutrino
synchrotron radiation, while in the present case the magnetic
field affects the bremsstrahlung process externally. We consider
the neutrino bremsstrahlung process in presence as well as in
absence of a magnetic field as two independent processes \footnote
{This result is published in {\it Journal of Physics G}
\cite{Bhattacharyya4}.}, and intend to calculate the scattering
cross-section and energy loss rate for both of them. A comparative
study can also be carried out to see how a magnetic field
influences the rate of neutrino-antineutrino emission. It is worth
mentioning that the screening effect is taken into account for the
process in absence of a magnetic field; but in presence of a
magnetic field such screening effect is no longer effective.
\subsection{Calculation of scattering cross-section}
The neutrino bremsstrahlung process occurs when an electron
collides with a nucleus to emit neutrino-antineutrino pair. If
there exists a strong magnetic field it will affect the momentum
of the electron and so will influences the bremsstrahlung process.
In that case the motion of the electron is dissimilar compared to
the situation when there is no magnetic field and thus the basic
structures for these two processes will be different. Figure-11
and Figure-12 represent the Feynman diagrams for neutrino
bremsstrahlung in presence of a magnetic field. In our
calculations the presence of a magnetic field plays a crucial role
and as such it is to be handled with care. Without violating the
generality we can choose $z$-axis along the direction of magnetic
field. The component of electron momentum along the direction of
magnetic field remains unaffected. It is clear that the effect of
the magnetic field on the electron quantizes its energy to the
direction perpendicular to $H$ and thus transverse components get
replaced by
$p_{x}^{2}+p_{y}^{2}\longrightarrow2nm_{e}^{2}\frac{H}{H_{c}}$,
whereas the longitudinal component $p_{z}$ is directed along the
magnetic field. The Feynman diagrams for the neutrino
bremsstrahlung process are almost similar for both in presence and
in absence of a magnetic field; only we have to keep in mind that
four momenta of the electronic lines in the diagrams, are to be
modified when a magnetic field is present. In presence of a
magnetic field the energy momentum relation of the electron is
modified to
$$(p_{n}^{0})^{2}=m_{e}^{2}+ p_{z}^{2}+2n\frac{H}{H_{c}}m_{e}^{2}\eqno{(4.2.1)}$$
where $H_{c}=4.414\times 10^{13}$ G stands for the critical
magnetic field and $n$ represents the Landau level for the
electron in the magnetic field. It is to be mentioned that the
twice averaged value of the fermion spin projection $s$ is related
to the Landau level $n$ by the following relation.
$$n=\nu+(1-\frac{e}{\mid e\mid}s)/2$$
where $\nu=0,1,2.....$. Clearly, $s=1$ if the electron spin is
along the magnetic field direction, and $s=-1$ if it is opposite
to the magnetic field direction. For the ground Landau level the
value of $s$ is taken as 1.
\\\indent Let us consider and calculate the bremsstrahlung process
in absence of a magnetic field. In this case the coulomb field has
the potential as follows.
$$\Phi(\overrightarrow{r})=\frac{Ze}{\mid\overrightarrow{r}\mid}
e^{-\mid\overrightarrow{r}\mid /\lambda_{d}}\eqno{(4.2.2)}$$ i.e.,
a single charge with an exponential screening cloud (Yukawa like
charge distribution), where $\lambda_{d}$ is the Debye screening
length given by
$$\frac{1}{\lambda_{d}^{2}}=\frac{4e^{2}}{\pi}E_{F}P_{F}\eqno{(4.2.3)}$$
$P_{F}$ and $E_{F}$ represent the Fermi momentum and energy,
respectively. We consider the potential, given by (4.2.2), because
the screening effect is important in the high density region. The
matrix element is constructed as
$$M_{fi}=-ie\frac{G_{F}}{\sqrt{2}}A_{0}(\overrightarrow{k})J_{\mu}\mathcal{M}^{\mu}
\eqno{(4.2.4)}$$ where,
$$\mathcal{M}^{\mu}=\hspace{1cm}\overline{u}(p')[\gamma^{\mu}(C_{V}-C_{A}\gamma_{5})
\frac{(p^{'\tau}\gamma_{\tau}+q^{\tau}\gamma_{\tau}+m_{e})}
{(p'+q)^{2}-m_{e}^{2}+i\epsilon}\gamma^{0}$$$$\hspace{5cm}+\hspace{0.2cm}\gamma^{0}
\frac{(p^{\tau}\gamma_{\tau}-q^{\tau}\gamma_{\tau}+m_{e})}
{(p-q)^{2}-m_{e}^{2}+i\epsilon}\gamma^{\mu}(C_{V}-C_{A}\gamma_{5})]u(p)\eqno{(4.2.5)}$$
$$J_{\mu}=\overline{u}_{\nu}(q_{1})\gamma_{\mu}(1-\gamma_{5})v_{\nu}(q_{2})\eqno{(4.2.6)}$$
$$A_{0}(\overrightarrow{k})=\int\Phi(\overrightarrow{r})
e^{-(\overrightarrow{k}.\overrightarrow{r})}d^{3}r=-\frac{4\pi
Ze}{\mid k^{2}+q_{sc}^{2}\mid}\eqno{(4.2.7)}$$ and
$$q=q_{1}+q_{2}$$ The energy momentum conservation gives
$$k+p=p'+q$$ where $k$ is purely space-like as in the case of photo-coulomb
neutrino process \cite{Rosenberg, Bhattacharyya1}. The term
$q_{sc}$, present in the equation (4.2.7), arises due to the
screening effect and can be expressed as
$$q_{sc}=\frac{1}{\lambda_{d}}\eqno{(4.2.8)}$$
Using some simplifications we can express the term
$J_{\mu}\mathcal{M}^{\mu}$ in the following way.
$$J_{\mu}\mathcal{M}^{\mu}=[\frac{(p'J)}{q^{2}/2+(p'q)}+\frac{(pJ)}{q^{2}/2-(pq)}]
\overline{u}(p')\gamma^{0}(C_{V}-C_{A}\gamma_{5})u(p)\eqno{(4.2.9)}$$
We put this expression to the equation (4.2.4) to obtain the
expression of the scattering matrix. It is sufficient to calculate
the expression
$\sum\mid\overline{u}(p')\gamma^{0}(C_{V}-C_{A}\gamma_{5})u(p)\mid^{2}$.
Thus we obtain
$$\sum\mid\overline{u}(p')\gamma^{0}(C_{V}-C_{A}\gamma_{5})u(p)\mid^{2}=(C_{V}^{2}-C_{A}^{2})
+(C_{V}^{2}+C_{A}^{2})\frac{(p^{'0}p^{0}-\overrightarrow{p}'.\overrightarrow{p})}
{m_{e}^2}\eqno{(4.2.10)}$$ To be noted that in the equation
(4.2.10) there is no term of the neutrino momentum, so this part
is not considered when we integrate over the final momenta of
neutrinos. Let us now evaluate the squared spin sum of the
expression
$\mid\frac{(p'J)}{q^{2}/2+(p'q)}+\frac{(pJ)}{q^{2}/2-(pq)}\mid$
and then integrating over final momenta of the neutrinos we obtain
[Appendix-A]
$$\int\sum\mid\frac{(p'J)}{q^{2}/2+(p'q)}+\frac{(pJ)}{q^{2}/2-(pq)}\mid^{2}
\frac{d^{3}q_{1}d^{3}q_{2}}{(2\pi)^{3}2q^{0}_{1}(2\pi)^{3}2q^{0}_{2}}(2\pi)
\delta(q^{0}-q^{0}_{1}-q^{0}_{2})$$
$$\approx\frac{1}{18(2\pi)^{3}m_{\nu}^{2}}(p^{0}-p^{'0})^{3}
\frac{\mid\overrightarrow{p}-\overrightarrow{p}'\mid^{2}}{(p^{0}+p^{'0})^{2}}\eqno{(4.2.11)}$$
Up to this step the calculations for the neutrino bremsstrahlung
process is identical in both situations $-$ in presence as well as
in absence of a magnetic field. It is assumed that in presence of
a magnetic field the quantized transverse components of the
momentum do not participate directly during the interaction
between nucleus and electron. Only the $z$-component of the
electron momentum takes part in this process. Thus in this case we
obtain an expression almost similar to the equation (4.2.11)
replacing $\mid\overrightarrow{p}'\mid$ and
$\mid\overrightarrow{p}\mid$ by $p'_{z}$ and $p_{z}$ respectively.
Now, to integrate the squared sum of the matrix element over all
final momenta we shall utilize the result obtained in the equation
(4.2.10) and (4.2.11). We are to remember that there is no
screening effect in presence of a strong magnetic field. In
presence of a magnetic field the phase space factor takes the form
[Appendix-B]
$$\int d^{3}p'=\pi \frac{H}{H_{c}}m_{e}^{2}\int dp_{z}'\eqno{(4.2.12)}$$
whereas in the ordinary neutrino bremsstrahlung process the
integration over the final momentum of electron is done in the
usual manner. In the extreme-relativistic limit we evaluate the
integral over the final momentum of electron and obtain the
following expression.
$$\int\sum\mid M_{fi}\mid^{2}\frac{d^{3}q_{1}d^{3}q_{2}d^{3}p'}
{(2\pi)^{3}2q^{0}_{1}(2\pi)^{3}2q^{0}_{2}(2\pi)^{3}2p^{'0}}
(2\pi)\delta(q^{0}-q^{0}_{1}-q^{0}_{2})$$
$$=\hspace{0.5cm}\frac{8G_{F}^{2}\alpha^{2}Z^{2}}{9(2\pi)^{3}(1+r^{2})^{2}}
(C_{V}^{2}+C_{A}^{2})\frac{(p^{0})^{3}}{m_{e}^{2}m_{\nu}^{2}}\eqno{(4.2.13)}$$
This expression is obtained for the neutrino bremsstrahlung
process in absence of a magnetic field. In presence of a magnetic
field this expression becomes
$$\frac{2G_{F}^{2}\alpha^{2}Z^{2}}{9(2\pi)^{3}}(C_{V}^{2}+C_{A}^{2})\frac{p^{0}}
{m_{\nu}^{2}}(\frac{H}{H_{c}})\eqno{(4.2.14)}$$ The term $r$
arises in the equation (4.2.13) due to the weak screening effect.
It is given by
$$r\approx \frac{q_{sc}}{p^{0}}$$
 The terms $C_{V}$ and $C_{A}$ for the electron type of
neutrino emission differ from those in case of muon and tau
neutrino emissions, since $W$-boson exchange diagrams are present
only when the electron neutrino-antineutrino pair is emitted. Now,
inserting the above two expressions [from the equations (4.2.13)
and (4.2.14)] into the expression of the scattering cross-section
for both the cases and switching over to the C.G.S. unit we
finally obtain
$$\sigma\approx1.76\times10^{-50}(\frac{E}{m_{e}c^{2}})^{2}\frac{1}{(1+r^{2})^{2}}
\hspace{0.5cm}cm^{2}\eqno{(4.2.15)}$$ in absence of a magnetic
field, whereas
$$\sigma_{mag}\approx4.41\times10^{-51}(\frac{H}{H_{c}})
\hspace{0.5cm}cm^{2}\eqno{(4.2.16)}$$ in presence of a magnetic
field.
\\ To be noted that in our calculations we have considered all the three types of
neutrino as earlier cases. Our result (equation 4.2.16) shows that
the scattering cross-section for the neutrino bremsstrahlung
process in presence of a magnetic field depends not on the energy
of the incoming electron, but on the intensity of a magnetic field
present.
\subsection{Calculation of energy loss rate}
In the extreme-relativistic case the energy loss rate in erg per
nucleus per second for the neutrino bremsstrahlung process is
calculated by the formula
$$\mathcal{E}_{\nu}^{Z}=\frac{2}{(2\pi)^{3}\hbar^{3}}
\int\frac{d^{3}p}{[e^{\frac{E-E_{F}}{\kappa T}}+1]}c\sigma E
e^{\frac{E-E_{F}}{\kappa T}}\eqno{(4.3.1)}$$ where $E_{F}$ stands
for the Fermi energy of the electron. We are considering the
situation where the electrons are highly degenerate. We know that
in the degenerate region the energy of the electron remains below
the Fermi energy level. To obtain the energy loss rate in erg per
gram per second, $\mathcal{E}_{\nu}^{Z}$ is divided by $Am_{p}$
which comes to
$$\mathcal{E_{\nu}}=\frac{Z^{2}}{A}\times5. 26\times10^{-3}\times
\frac{x_{F}^{6}e^{1-x_{F}}} {(1+r^{2})^{2}}T_{10}^{6}\hspace{0.5
cm}erg/gm-sec\eqno{(4.3.2)}$$ where, $T_{10}=T\times
10^{-10}$\\The $x_{F}$ represents the ratio of the Fermi
temperature to the maximum temperature of the degenerate electron
gas. The degeneracy is attained only when the following condition
is satisfied \cite{Chandrasekhar}.
$$x_{F}^{2}>2\pi^{2}\eqno{(4.3.3)}$$
We have calculated $x_{F}\approx 6$, considering the fact that the
temperature and density of the electron gas present in the core of
a newly born neutron star would be approximately $10^{12}$ K and
$10^{15}$ $gm/cc$, respectively. The term $r$, arising due to the
weak screening in absence of a magnetic field, is related to
$x_{F}$ by $r\approx 0. 096 \times x_{F}$. Thus we can calculate
the term $r$. Finally, the expression for the energy loss rate in
absence of a magnetic field is obtained as
$$\mathcal{E_{\nu}}=\frac{Z^{2}}{A}\times0.93\times10^{12}\times
T_{10}^{6} \hspace{0.5 cm}erg/gm-sec\eqno{(4.3.4)}$$ In the same
manner we obtain the energy loss rate in presence of a magnetic
field. In that case the phase space factor is replaced according
to the rule defined in (4.2.12). Thus we calculate the energy loss
rate in presence of a magnetic field as
$$\mathcal{E_{\nu}}^{mag}=\frac{Z^{2}}{A}\times0.51\times10^{6}
\times
H_{13}^{2}T_{10}^{2}\hspace{0.5cm}erg/gm-sec\eqno{(4.3.5)}$$
where, $H_{13}=H\times 10^{-13}$ \\We have computed (Table-5)
the logarithmic value of the energy loss rate in the temperature
range $0. 8\times 10^{10} - 10^{11}$ K and the magnetic field
$10^{14}- 10^{16}$ G at a fixed density $\rho=10^{15}$ $gm/cc$.
\subsection{Discussion}
The neutrino bremsstrahlung process is an important mechanism for
ascertaining the energy loss in the stellar core during the
stellar evolution. This process may be very effective in the
highly degenerate region, for example, the core of the low mass
red giant, white dwarf etc. Along with this degenerate nature if
the electron gas is highly relativistic, the energy loss rate
through the bremsstrahlung process becomes significantly high.
Therefore, it may be a significant source of energy loss through
neutrino-antineutrino emission during neutron star cooling which
characterizes a highly relativistic degenerate region. It was
calculated that the neutrino luminosity in the crust of a neutron
star is high enough, but it is to be verified what will be the
effect of neutrino emission by the bremsstrahlung process in the
core region, particularly when the core is strongly magnetized. We
know in the neutron stars and magnetars the magnetic field may
reach to $10^{16}$ G and may influence the neutrino bremsstrahlung
process. We are, here, verifying how a magnetic field affects the
bremsstrahlung process $-$ whether it increases the rate of energy
loss or decreases it. In the neutrino synchrotron radiation,
neutrino-antineutrino pair emission takes place since the electron
changes its Landau levels, but in the bremsstrahlung the Landau
levels are assumed to remain unchanged throughout the process. The
neutrino bremsstrahlung process occurs through the change of
magnitude of the longitudinal component of the electron linear
momentum directed along the magnetic field.\\\indent Our
calculations show that the presence of a strong magnetic field
weakens the neutrino bremsstrahlung process in the early stage of
neutron star cooling resulting the decrease of energy loss rate,
whereas in some stages it (the strong magnetic field) facilitates
the process. This is clear if we study the Table-5 where the
energy loss rate is computed in the logarithmic scale. It is
observed from this table that in the temperature range $10^{10}$ K
$\leq T\leq 10^{11}$ K and at the density $10^{15}$ $gm/cc$ the
energy loss rate for the neutrino bremsstrahlung process in
absence of a magnetic field is greater than that in presence of a
strong magnetic field ($10^{14} - 10^{16}$ G). We interpret that
in the early stage of neutron star cooling, when the temperature
remains above $10^{10}$ K, the effect of the neutrino
bremsstrahlung process gets lowered due to the presence of a
strong magnetic field, although the energy loss rate is still very
high. If the strength of the magnetic field goes well below the
critical value, the process becomes free from the influence of
this magnetic field, and a greater amount of energy is lost. If we
look at the table more carefully then it appears that there exists
a particular region ($5.93\times 10^{9}$ K $<T< 10^{10}$ K, $H\sim
10^{16}$ G and $\rho\sim 10^{15}$ $gm/cc$) \footnote {More
specifically the temperature range is $5.93\times 10^{9}$ K $<T<
9\times 10^{10}$ K.} during a neutron star cooling where the
neutrino bremsstrahlung process contributes a greater amount of
energy loss by the influence of a magnetic field compared to the
situation when there is no magnetic field at all. This is an
important consequence of our study that in a particular region the
magnetic field makes the neutrino bremsstrahlung process more
rapid, but in general the process becomes less effective in
presence of a magnetic field. Therefore, it is ascertained that if
the temperature falls below $10^{10}$ K, the process gives maximum
effect due to the presence of a strong magnetic field. We
emphasize the point that the process is accelerated by a magnetic
field when its intensity reaches $10^{16}$ G, i.e., it is possible
only in case of neutron stars and magnetars having a very high
magnetic field. Our study reveals that the neutrino bremsstrahlung
process has an important contribution to the energy loss in the
highly degenerate stellar core during the late stages of stellar
evolution in presence of a strong magnetic field.
\subsection{Appendix-A} We have chosen a frame in which
$$\overrightarrow{q}=\overrightarrow{q}_{1}+\overrightarrow{q}_{2}=0\eqno{(A.1)}$$
In this frame we obtain
$$\frac{q^{2}}{2}+(p'q)=-\frac{(p^{'0}-p^{0})(p^{'0}+p^{0})}{2}\eqno{(A.2)}$$
$$\frac{q^{2}}{2}-(pq)=\frac{(p^{'0}-p^{0})(p^{'0}+p^{0})}{2}\eqno{(A.3)}$$
Thus,
$$\sum\mid\frac{(p'J)}{q^{2}/2+(p'q)}+\frac{(pJ)}{q^{2}/2-(pq)}\mid^{2}
=\frac{4}{(p^{'0}-p^{0})^{2}(p^{'0}+p^{0})^{2}}\sum\mid(p'-p)J\mid^{2}\eqno{(A.4)}$$
Let us take
$$P=p-p'=q-k\eqno{(A.5)}$$
\vspace{0.5cm} We have,
$$\sum\mid(PJ)\mid^{2}=\sum\mid\overline{u}_{\nu}(q_{1})P^{\mu}\gamma_{\mu}(1-\gamma_{5})
v(q_{2})\mid^{2}$$
$$\hspace{1.5cm}=\frac{2}{m_{\nu}^{2}}[2(q_{1}P)(q_{2}P)-(q_{1}q_{2})P^{2}]$$
$$\hspace{5cm}=\frac{2}{m_{\nu}^{2}}[(m_{\nu}P^{0})^{2}+\{(1-2\cos^{2}\alpha)
\mid\overrightarrow{q}_{1}\mid^{2}+(q_{1}^{0})^{2}\}\mid
\overrightarrow{P}\mid^{2}]\eqno{(A.6)}$$ Now using
$$\int d^{3}q_{2}=\frac{4\pi}{3}\mid\overrightarrow{q}_{2}\mid^{3}
=\frac{4\pi}{3}\mid\overrightarrow{q}_{1}\mid^{3}\eqno{(A.7)}$$
and
$$d^{3}q_{1}=\mid\overrightarrow{q}_{1}\mid^{2}
d\mid\overrightarrow{q}_{1}\mid \sin\alpha\hspace{0.05cm}
d\alpha\hspace{0.05cm} d\phi\eqno{(A.8)}$$ we obtain
$$\int\sum\mid (PJ)\mid^{2}\frac{d^{3}q_{1}}{2q_{1}^{0}}\frac{d^{3}q_{2}}{2q_{2}^{0}}
\delta(2q_{1}^{0}-q^{0})$$
$$=\frac{2\pi^{2}}{3m_{\nu}^{2}}\int\int_{\alpha=0}^{\pi}
\frac{\mid\overrightarrow{q}_{2}\mid^{4}}{q_{1}^{0}}[(m_{\nu}P^{0})^{2}+\{(1-2\cos^{2}\alpha)
\mid\overrightarrow{q}_{1}\mid^{2}+(q_{1}^{0})^{2}\}\mid
\overrightarrow{P}\mid^{2}]$$$$\delta(q_{1}^{0}-\frac{q^{0}}{2})dq_{1}^{0}
\hspace{0.05cm}\sin\alpha\hspace{0.05cm} d\alpha\hspace{0.05cm}
d\phi$$
$$=\frac{4\pi^{2}}{3m_{\nu}^{2}}\int\frac{\mid\overrightarrow{q}_{2}\mid^{4}}{q_{1}^{0}}
[(m_{\nu}P^{0})^{2}+\{\frac{\mid\overrightarrow{q}_{1}\mid^{2}}{3}+(q_{1}^{0})^{2}\}\mid
\overrightarrow{P}\mid^{2}]\delta(q_{1}^{0}-\frac{q^{0}}{2})dq_{1}^{0}$$
$$\approx \frac{\pi^{2}}{18m_{\nu}^{2}}(p^{0}-p^{'0})^{5}\mid
\overrightarrow{p}-\overrightarrow{p}'\mid^{2}\eqno{(A.9)}$$ We
have assumed $m_{\nu}\ll q_{1}^{0}<q^{0}$ and used the following
criteria:
$$m_{\nu}\longrightarrow 0$$
$$P^{0}=q^{0}=p^{0}-p^{'0}$$
$$\overrightarrow{P}=\overrightarrow{k}=\overrightarrow{p}-\overrightarrow{p}'$$
since $k$ is space-like, whereas $q$ is time-like in our chosen
frame.\\\indent Now introducing normalized factors and also using
$(A.4)$ we obtain
$$\int\sum\mid\frac{(p'J)}{q^{2}/2+(p'q)}+\frac{(pJ)}{q^{2}/2-(pq)}\mid^{2}
\frac{d^{3}q_{1}d^{3}q_{2}}{(2\pi)^{3}2q^{0}_{1}(2\pi)^{3}2q^{0}_{2}}(2\pi)
\delta(q^{0}-q^{0}_{1}-q^{0}_{2})$$
$$\approx\frac{1}{18(2\pi)^{3}m_{\nu}^{2}}(p^{0}-p^{'0})^{3}\frac{\mid\overrightarrow{p}
-\overrightarrow{p}'\mid^{2}}{(p^{0}+p^{'0})^{2}}\eqno{(A.10)}$$
This is same as the equation (4.2.11).
\subsection{Appendix-B} In presence of a magnetic field the phase space
factor is replaced by the following relation \cite{Roulet}
$$\frac{2}{(2\pi)^{3}}\int d^{3}p=\frac{1}{(2\pi)^{2}}\sum_{n=0}^{n_{max}}g_{n}
\int dp_{z}\eqno{(B.1)}$$ where $g_{n}$ represents degeneracy
factor of the Landau levels, i.e.,
$$g_{0}=1,\hspace{4cm} g_{n}=2\hspace{2cm}(n\geq 1)\eqno{(B.2)}$$
The maximum Landau level $n_{max}$ is obtained from the following
relation
$$n_{max}=\frac{1}{2m_{e}^{2}}(\frac{H}{H_{c}})[(p^{0}_{n_{max}})^{2}-(p^{0})^{2}]\eqno{(B.3)}$$
where,
$$(p^{0})^{2}=p_{z}^{2}+m_{e}^{2}\eqno{(B.4)}$$
For $n_{max}<1$ we have,
$$H > \frac{1}{2m_{e}^{2}}[(p^{0}_{n_{max}})^{2}-(p^{0})^{2}]H_{c}\eqno{(B.5)}$$
It shows that for a high magnetic field only $n=0$ Landau level
contributes in the phase space. In this article we consider the
stellar region is highly magnetized. It gives
$$(p^{0}_{n_{max}})^{2}-(p^{0})^{2}> 2m_{e}^{2}$$
and therefore,
$$H > H_{c}$$
In that case from (B.1) we obtain
$$\int d^{3}p=\pi \frac{H}{H_{c}}m_{e}^{2}\int dp_{z}\eqno{(B.6)}$$
It is same as the equation (4.2.12).\\\indent If the magnetic
field is comparatively lower the higher Landau levels contribute
in the phase space as per the condition (B.3).

\section{Some more important processes}
Now we shall discuss some neutrino emission processes which may
play important role during the late stages of stellar evolution,
other than the four we have considered in this paper.

\subsection{Plasma neutrino process}
The decay of plasmons $(\Gamma\rightarrow\nu+\overline{\nu})$,
termed as plasma neutrino process, into neutrino-antineutrino pair
is considered to be very effective mechanism to bring out energy
from the stellar core, particularly in the degenerate region. In
the plasma the vibrations of both electromagnetic waves and the
charged particles generate transverse as well as longitudinal
waves. When the dielectric constant of the electron gas becomes
less than unity the photon behaves as if it has a rest mass that
is equal to the plasma frequency. If the plasma frequency
($\hbar\omega_{p}$) is not negligible relative to the temperature
($\kappa T$) of the stellar core, the collective behavior of the
plasmon is more significant than the effect from a single photon
and electron. In 1963 Adams et al. \cite{Adams} considered the
plasma decay process to carry out the calculations in
astrophysical scenario. They calculated the energy loss rate for
the decay of both transverse and longitudinal plasma in the low as
well as high temperature limit. They also indicated that the
neutrino luminosity of a star could far exceed its photon
luminosity at a central temperature higher than $10^{8}$ K, and
also at the higher density where it would have dominant effect
compared to the pair annihilation process resulting in neutrino
pair emission. In 1972 Dicus \cite{Dicus1972} considered various
neutrino emission processes in the electro-weak theory. In that
paper he outlined the plasma neutrino process and briefly
discussed its importance. In 1990 Braaten \cite{Braaten1} modified
the dispersion relation used in the earlier calculations. In the
high temperature limit, the use of the new ultra-relativistic
dispersion relation would increase the emissivity by a factor of
3.185. In 1991 Braaten and Segel \cite{Braaten2} carried out a
detailed calculations related to the plasma neutrino process. They
deduced a generalized dispersion relation and obtained a bit
modified expression of the neutrino emissivity for the decay of
plasma in the transverse, longitudinal and axial-vector modes.
Their result indicated that the longitudinal emissivity would be
suppressed relative to the transverse emissivity by a factor of
$\frac{\hbar\omega_{p}}{\kappa T}$ in the high-temperature limit.
The decay of plasma into neutrino-antineutrino pair is thus a
widely discussed topic and believed to be one of the very
important process in the stages where the neutrino emission
dominates over the electromagnetic radiation.\\\indent The plasma
neutrino process is also important in presence of a magnetic
field. Canuto et al. \cite{Canuto2} considered plasma neutrino
process in presence of a strong magnetic field ($10^{12}-10^{13}$
G). They studied two cases in which the propagations are parallel
and perpendicular to the magnetic field. In the first paper
\cite{Canuto2} they considered the effect of an intense magnetic
field on the decay of transverse plasmon into
neutrino-antineutrino pair. It was found that the effect would be
negligible in the regions of astrophysical interest since the
cyclotron frequency
$\omega_{c}=\frac{H}{H_{c}}\frac{mc^{2}}{\hbar}$ is very small in
comparison with the plasma frequency $\omega_{p}$. In the
subsequent paper \cite{Canuto2} they studied the decay of
longitudinal plasmon in presence of the strong magnetic field.
Contrary to the transverse case, the decay of longitudinal
plasmons would have a significant contribution in presence of a
magnetic field. In 1998 Kennett and Melrose \cite{Kennett} carried
out detailed calculations of such anisotropic plasma decay into
$\nu\overline{\nu}$ pair. They calculated the response tensor in
presence of a strong magnetic field for cold as well as thermal
plasma. They introduced a vertex function to include the axial
vector current of the weak interaction and showed that contrary to
the case of unmagnetized plasma the axial vector coupling could
have a role in affecting the neutrino emission via plasma process.
They also suggested a criterion with which to estimate such axial
vector effects that would be important. Hence, the decay of plasma
is not only highly significant in absence of a magnetic field, but
in presence of a magnetic field it may have an important
contribution in the late stages of stellar evolution.

\subsection{Photo-neutrino process}
The photo-neutrino process ($\gamma+e^{-}\rightarrow
e^{-}+\nu+\overline{\nu}$) is one of the significant neutrino
emission processes carrying away energy from the evolved star. In
1961 Chiu and Stabler \cite{Chiu1961} considered this process and
calculated the energy loss rate (in $erg/gm-sec$) for
non-degenerate as well as degenerate electron. Dicus recalculated
\cite{Dicus1972} the photo-neutrino process in the framework of
the electro-weak theory in which neutrino pair production takes
place through the exchange of intermediate Z and W bosons. In his
calculation Dicus included the muon neutrino (but not the tau
neutrino) and obtained a modified result. Dutta et al.
\cite{Dutta} derived the neutrino emissivity for the
photo-neutrino process in hot and dense matter, although here
instead of ordinary photon the massive photon (plasmon) was
considered. They presented numerical results for widely varying
conditions of temperature and density. As regards ordinary photon
the photo-neutrino process is supposed to play an important role
for the low density $\frac{\rho}{\mu_{e}}\leq 10^{5}$ $gm/cc$ and
comparatively low temperature $T\leq 4\times 10^{8}$ K.

\subsection{Pair-annihilation process}
Another important process having significant contribution to the
energy loss in the stellar core is pair-annihilation process ($
e^{+}+e^{-}\rightarrow\nu+\overline{\nu}$). In 1960 Chiu and
Morrison \cite{Chiu1960} for the first time considered the
pair-annihilation process. After that Chiu recalculated the
process in the paper \cite{Chiu1961} with Stabler and obtained the
energy loss rate $\mathcal{E}^{pair-annihilation}_{\nu}$ in
different regions. Dicus \cite{Dicus1972} carried out detailed
calculations of pair-annihilation process in different regions.
That was an extensive work. He calculated the energy loss rate for
this process as well as other important neutrino emission
processes. The pair-annihilation process is important in the
non-degenerate case, especially when the electron gas becomes
relativistic. Undoubtedly, the process has such a dominant role in
this region that it is considered to be an important mechanism
during the stellar evolution.\vspace{1cm}\\All these neutrino
emission processes, discussed in this section, are already known
to be important processes occurring in the stellar core. It is
found that the combination of such neutrino emission processes is
the main source of energy loss during the later phases of the
evolution of stars. In the following section we have discussed
different regions which are relevant during the stellar evolution.
\section{Some important stellar regions} In the later stages of
the stellar evolution the stellar objects can be classified into
four major regions depending upon the density and temperature of
the electron gas present in their cores. It is obvious that the
temperature of the stellar core during the later stages of the
evolution of stars is nearer to or even more than $10^{8}$ K. In
some cases like white dwarf cores the temperature may be around
$10^{7}$ K. In the highly relativistic and degenerate stellar
objects, such as neutron stars, the core temperature may be higher
than $10^{11}$ K. We have considered such regions in our study
since the neutrino emissions depend on the temperature and density
(and sometime on magnetic field). We have got an impression from
the study of different neutrino emission processes that all
mechanisms may not be effective throughout the entire stages of
stellar evolution, rather in a specific region a particular
process may have a significant effect. Let us first brief the four
main regions \cite{Chiubook}.
\begin{enumerate}
\item {\bf Non-relativistic non-degenerate region:} This region is
characterized by the temperature $T\leq 5.93 \times 10^{9}$ K and
$(\frac{\rho}{\mu_{e}})^{2/3}\leq \frac{T}{2.97\times 10^{5} K}$.
The important stellar objects in this region are massive stars
beyond helium burning stage, nuclei of the planetary nebula and
possibly quasars. \item {\bf Non-relativistic degenerate region:}
This region is confined within the range $T\leq 5.93 \times
10^{9}$ K, $\rho\leq 10^{6}$ $gm/cc$ and
$(\frac{\rho}{\mu_{e}})^{2/3}\geq \frac{T}{2.97\times 10^{5} K}$.
Along with the massive stars beyond helium burning stage other
stellar objects in this region are white dwarf, pre-white dwarf
stars and population-II red giant. \item {\bf Relativistic
non-degenerate region:} This region is observed in the stellar
bodies when the electron gas, present within them, becomes
relativistic in nature.  The physical conditions required to
attain this region are $T\geq 5.93 \times 10^{9}$ K and
$\frac{\rho}{\mu_{e}}\leq (\frac{T}{5.93\times 10^{7} K})^{3}$.
Pre-supernova stars belong to this region. The stellar collapse
too is an example of the relativistic non-degenerate region. \item
{\bf Relativistic degenerate region:} This is an important region
related to the end stage of the stellar evolution and most of the
neutrino emission processes we have studied in this paper, are
relevant in this region. This region is characterized by the
physical conditions $T\geq 5.93 \times 10^{9}$ K and
$\frac{\rho}{\mu_{e}}\geq (\frac{T}{5.93\times 10^{7} K})^{3}$.
The main stellar object in this region is the neutron star,
especially in its newly born stage. The magnetar is another
stellar object in the relativistic degenerate region.
\vspace{0.02cm}
\end{enumerate}

Besides these four regions there is another factor which plays an
important role in some neutrino emission processes $-$ the
presence of strong magnetic field. Although we know very little
about the magnetic field inside the stars, it is possible that
some of the degenerate stars may have very large internal magnetic
field. A number of stars on or near the main sequence,
particularly A-type stars, have been observed to have surface
magnetic field of the order of $10^{3} - 10^{4}$ G, which suggests
that there are fields of the same order of magnitude in the
interior of these stars. It has already been indicated in Chapter
3 that in some white dwarves, which are the end stages of low mass
stars, the magnetic field may reach up to $10^{11}$ G. It has also
been mentioned that the neutron star, which is extremely
relativistic and highly degenerate, may have a strong magnetic
field. In some neutron stars the magnetic field, if present, may
be very high ($10^{12}-10^{16}$ G). The neutron star having a very
high magnetic field is called a magnetar \cite{Duncan}. When in a
supernova a star collapses to a neutron star, its magnetic field
increases drastically in strength. It is estimated that about 1 in
10 supernova explosions results in a magnetar instead of a
standard neutron star or pulsar. In the following sections we
discuss the effect of neutrino emission processes in different
regions in both absence and presence of a strong magnetic field.

\section{Region of importance of different processes in absence of magnetic field}
Most of the stellar objects during their evolution period are
non-magnetic. In the later stages of the stellar evolution, when
there is no magnetic field, the photo-neutrino, pair annihilation,
plasma neutrino and neutrino bremsstrahlung are considered to have
significant contributions to carry away the energy through the
emission of neutrinos. The details are in the book written by Chiu
\cite{Chiubook} and later the most updated considerations of the
above processes are discussed in the book written by Raffelt
\cite{Raffelt}. It is found that in the comparatively low
temperature and density range the photo-neutrino process is much
effective and dominating over all other processes. Thus the
photo-neutrino is the dominating process in the non-relativistic
non-degenerate region. The scenario changes with the increasing
temperature in the same density range characterizing the
relativistic non-degenerate region. In this region the pair
annihilation process becomes the main source of energy loss. In
the degenerate region the plasma neutrino process plays a dominant
role over other neutrino emission processes. It is to be noted
that although the neutrino bremsstrahlung has a significant
contribution in the relativistic degenerate region, the plasma
neutrino process has the greater effect. A diagram is provided in
Chiu's book \cite{Chiubook} to show in which region which process
has the dominant role. It clearly indicates the region of
importance of different processes during the later stages of the
stellar evolution. This diagram was drawn from the results
obtained from current current coupling theory. The diagram is also
consistent with the result obtained by Dicus \cite{Dicus1972},
although he calculated the process according to the electro-weak
theory. In this paper we have studied two important non-magnetic
processes. We would like to study in which region those processes
have important effect. It is to be pointed out that Dicus
\cite{Dicus1972} considered two types of neutrino in his
calculations, whereas we consider all three types. We include tau
neutrino in our calculations. We have calculated all processes
(non-magnetic and magnetic) strictly following the Standard Model;
only a minor relaxation is allowed when we consider a small
neutrino mass in our calculations.\\\indent At first, we have
considered the photo-coulomb neutrino process and found this
process to have some effect in the non-degenerate region,
especially at the temperature close to $10^{9}$ K. In the
non-relativistic and non-degenerate region the most dominating
process is the photo-neutrino process, but with the high
temperature and low density (relativistic non-degenerate region)
pair annihilation process is more dominating. The photo-coulomb
neutrino process at the temperature $10^{9}$ K in this region with
its maximum possible effect has the lesser contribution compared
to the above two processes. Thus the photo-coulomb neutrino
process is not the dominant mechanism in any particular region,
though its role cannot be ignored. A point is to be noted that the
analytical expression of energy loss rate for the photo-coulomb
neutrino process has a strong temperature dependence $(\sim
T^{10})$; therefore, the energy loss rate increases rapidly with
the increasing temperature. Thus in a very short range of
temperature this process has a significant contribution. Below the
temperature $5\times 10^{8}$ K this process practically has
insignificant contribution, but around $10^{9}$ K it suddenly
becomes significant.\\\indent Another important process
(non-magnetic) we have studied, is the electron-neutrino
bremsstrahlung. We have calculated this process in the framework
of the electro-weak theory in the different regions of the stellar
evolution in late stages. In the non-relativistic non-degenerate
region the electron-neutrino bremsstrahlung has a little effect.
In the range of temperature $10^{8}-10^{9}$ K and density
$10^{2}-10^{6}$ $gm/cc$ the contribution of this process cannot be
more than that of photo-neutrino process. The electron-neutrino
bremsstrahlung process has significant contribution in the
extreme-relativistic region. In the extreme-relativistic
non-degenerate region the dominating neutrino emission process is
pair annihilation. Its effect is so dominating that the
contribution of the electron-neutrino bremsstrahlung remains
suppressed, though the energy loss rate of the later process is
also very high in the relativistic non-degenerate region. In the
same manner, though the electron-neutrino bremsstrahlung
contributes a high energy loss rate in the extreme-relativistic
degenerate region, the plasma neutrino process has a greater
contribution in this region. Therefore, the electron-neutrino
bremsstrahlung process cannot be dominating in any particular
region, although its effect must be taken into account when we
consider the neutrino emission processes from stellar core. In
Figure-13 we have presented the region of dominance of different
neutrino emission processes in absence of magnetic field. Of
course, the photo-coulomb neutrino and electron-neutrino
bremsstrahlung have not been shown in the figure as none of these
is dominant in any particular region, but the contributions of
both of these processes are to be taken into account while
calculating the total energy loss rate by all neutrino emission
processes during the later stages of stellar evolution.
\section{Region of importance of different processes in presence of magnetic field}
The presence of a high magnetic field may influence the neutrino
emission rate during the evolution of stars. We have considered
two important processes which are relevant in presence of a strong
magnetic field. This kind of high magnetic field, close to or
greater than the critical magnetic field, is observed when the
stellar core is highly degenerate and relativistic in nature. It
is important to note that we have not calculated the plasma
neutrino process in presence of a strong magnetic field. In fact,
in absence of a magnetic field the decay of plasma is dominating
in the region indicated here (highly degenerate and extremely
relativistic). Also in presence of a strong magnetic field the
plasma neutrino process may have a significant contribution in
this region. However, in our study we are not considering this
effect. Here, we consider only the processes which have been
studied in this paper $-$ neutrino synchrotron radiation and
neutrino bremsstrahlung. The neutrino synchrotron process is a
widely discussed topic and found to have enormous effect in the
cooling of magnetized neutron stars. In addition to this we have
considered the effect of a strong magnetic field on the neutrino
bremsstrahlung process. We have found that this magnetic neutrino
bremsstrahlung process has also some remarkable effects during the
cooling of magnetar like stellar objects, particularly, in their
newly born stages. We have carried out a comparative study between
these two important magnetic processes at $H=10^{12}$ G in the
density range $10^{12}-10^{15}$ $gm/cc$ to find out the regions of
dominance. It comes out from our study that both the processes
have significant effects in the extreme-relativistic degenerate
region in presence of a strong magnetic field. Consequently, the
neutrino bremsstrahlung process is dominating at a comparatively
low temperature, although in the region under consideration
(extreme-relativistic and highly degenerate) the temperature is
much greater than $5.93\times10^{9}$ K. In the Figure-14 the
region of importance of these two processes are indicated at the
magnetic field $H=10^{12}$ G. Such a picture in presence of the
strong magnetic field is the result of our detailed study. Of
course, the regions of dominance change if we consider the
comparison at different magnetic field which may vary from
$10^{12}-10^{16}$ G in the magnetized neutron stars or magnetars.
Our study confirms that the above two processes are necessarily to
be considered for the energy loss in the star where the magnetic
field is very high.
\section{Concluding Remarks}
We have carried out a study concerning the role of some neutrino
emission processes as the stellar energy loss mechanisms. All
those processes may not be effective simultaneously, rather in a
specific region a particular process may have a remarkable effect.
In this paper we have calculated four neutrino emission processes
$-$ two in presence of strong magnetic field. In the preceding
sections we have discussed the role of different neutrino emission
processes in different regions during stellar evolution. It has
been pointed out that although the electron-neutrino
bremsstrahlung and photo-coulomb neutrino are not dominating
neutrino emission processes in any stage of stellar evolution, but
they may play significant role in some stages. Besides that, the
other two processes, we have studied in presence of a strong
magnetic field, are important when the stellar objects are
extremely magnetized ($10^{12}-10^{16}$ G). Finally, we have
indicated the region of dominance of different neutrino emission
processes both in absence and in presence of a magnetic field,
which gives an idea about the significance of different neutrino
emission processes in the different stellar regions during the
late stages of evolution of stars.\vspace{0.5cm}\\
{\Large \bf Acknowledgement:}\vspace{0.2cm}\\
I am highly grateful to Prof. Probhas Raychaudhuri, Ex-Professor, Department of Applied Mathematics, University of Calcutta for his valuable suggestions and unstinted support. I would also like to thank Mr. Asoke
Mukherjee for sparing his valuable time to go through the
manuscript and gave some useful suggestions. I must thank
{\bf Council of Scientific and Industrial Research}, India for
funding the research work.

\begin{table}
\begin{tabular}{|c|cc|}\hline
$T_{8}$&\hspace{3cm}$\frac{L_{\nu}}{L_{\odot}}$\\\hline & Current
current coupling th. & Electro-weak th.\\\hline
  1 & $2.21\times 10^{-11}$ & $6.07\times 10^{-11}$ \\
  2 & $7.68\times 10^{-9}$ & $2.11\times 10^{-8}$ \\
  3 & $8.1\times 10^{-7}$ & $2.22\times 10^{-6}$ \\
  4 & $2.32\times 10^{-5}$ & $6.36\times 10^{-5}$ \\
  5 & $3.17\times 10^{-4}$ & $8.68\times 10^{-4}$ \\
  6 & $2.79\times 10^{-3}$ & $7.65\times 10^{-3}$\\
  7 & $1.8\times 10^{-2}$ & $4.96\times 10^{-2}$ \\
  8 & $9.47\times 10^{-2}$ & $2.59\times 10^{-1}$ \\
  9 & $4.24\times 10^{-1}$ & $1.16$ \\
  10 & $5.8$ & $1.59\times 10$ \\\hline
\end{tabular}
\caption{Neutrino luminosity due to the photo-coulomb neutrino
process at a fixed density $ 10^{5}$ $gm/cc$ in the temperature
range $10^{8}- 10^{9}$ K}
\end{table}
\begin{table}
\begin{tabular}{|c|cc|}\hline
CM Energy $(MeV)$&\hspace{3cm}$\sigma_{\nu_{e}}(cm^{2})$\\\hline &
Software generated result  & Our result \\\hline
  10 & $3.45\times 10^{-48}$ & $3.64\times 10^{-48}$ \\
  20 & $1.19\times 10^{-47}$ & $1.88\times 10^{-47}$ \\
  30 & $4.47\times 10^{-47}$ & $4.84\times 10^{-47}$ \\
  40 & $1.17\times 10^{-46}$ & $0.92\times 10^{-46}$ \\
  50 & $1.23\times 10^{-46}$ & $1.56\times 10^{-46}$ \\
  60 & $1.95\times 10^{-46}$ & $2.32\times 10^{-46}$\\
  70 & $4.06\times 10^{-46}$ & $3.28\times 10^{-46}$ \\
  80 & $4.45\times 10^{-46}$ & $4.44\times 10^{-46}$ \\\hline
\end{tabular}
\caption{Comparison of the scattering cross-section for the
electron-neutrino bremsstrahlung process for electron type of
neutrino obtained by our calculation relative to that generated by
CalcHep software.}
\end{table}
\begin{table}
\begin{tabular}{|c|cc|}\hline
$T_{7}$&\hspace{3cm}$log\frac{L_{\nu}}{L_{\odot}}$\\\hline & Our
result. & Landstreet's result\\\hline
  $1.0$ & $-8.75$ & $-8.96$ \\
  $1.5$ & $-7.69$ & $-7.85$ \\
  $2.0$ & $-6.94$ & $-7.06$ \\
  $2.5$ & $-6.36$ & $-6.44$ \\
  $3.0$ & $-5.88$ & $-5.94$ \\
  $3.5$ & $-5.48$ & $-5.52$\\
  $4.0$ & $-5.13$ & $-5.15$ \\
  $4.5$ & $-4.83$ & $-4.83$ \\
  $5.0$ & $-4.55$ & $-4.54$ \\
  $5.5$ & $-4.3$ & $-4.28$ \\\hline
\end{tabular}
\caption{Neutrino luminosity (logarithmic scale) due to the
neutrino synchrotron radiation in a white dwarf ($\rho= 10^{8}$
 $gm/cc$ and $H=10^{11}$ G)}
\end{table}
\begin{table}
\begin{tabular}{|c|ccc|ccc|}\hline
$T_{9}$&&$log\frac{L_{\nu}}{L_{\odot}}$\\\hline & &Our result&&&
Landstreet's result&\\\hline & $10^{15}$ & $10^{14}$ & $10^{13}$ &
$10^{15}$ & $10^{14}$ & $10^{13}$\\\hline
  $6$ & $5.9$ & $6.55$ & $7.23$ & $6.51$ &
$7.06$ & $7.62$ \\
  $7$ & $6.30$ & $6.95$ & $7.64$ & $6.93$ &
$7.49$ & $8.04$  \\
  $8$ & $6.65$ & $7.3$ & $7.98$ & $7.3$ &
$7.85$ & $8.41$ \\
  $9$ & $6.96$ & $7.6$ & $8.29$ & $7.62$ &
$8.18$ & $8.73$ \\
  $10$ & $7.23$ & $7.88$ & $8.56$ & $7.91$ &
$8.47$ & $9.02$ \\\hline
\end{tabular}
\caption{Neutrino luminosity (logarithmic scale) due to the
neutrino synchrotron radiation in a neutron star ($\rho= 10^{15},
10^{14}, 10^{13}$ $gm/cc$ and $H=H_{c}=4.414\times 10^{13}$ G)}
\end{table}
\begin{table}
\begin{tabular}{|c|ccc|c|}\hline
$T_{10}$&&&$log(\frac{A}{Z^{2}}\mathcal{E_{\nu}})$\\\hline
&&&Presence of magnetic field& Absence of magnetic field\\\hline &
$10^{14}$ & $10^{15}$ & $10^{16}$\\\hline
$0.8$ & $7.51$ & $9.51$ & {\bf 11.51} & $11.39$ \\
$0.9$ & $7.62$ & $9.61$ & $11.62$ & $11.69$ \\
$1$ & $7.71$ & $9.71$ & $11.71$ & $11.97$ \\
$2$ & $8.31$ & $10.31$ & $12.31$ & $13.77$\\
$3$ & $8.66$ & $10.67$ & $12.67$ & $14.83$ \\
$4$ & $8.91$ & $10.91$ & $ 12.91$ & $15.58$ \\
$5$ & $9.10$ & $11.10$ & $13.10$ & $16.16$ \\
$6$ & $9.26$ & $11.26$ & $13.26$ & $16.64$ \\
$7$ & $9.40$ & $11.40$ & $ 13.40$ & $17.04$ \\
$8$ & $9.51$ & $11.51$ & $13.51$ & $17.39$\\
$9$ & $9.62$ & $11.62$ & $13.62$ & $17.69$ \\
$10$ & $9.71$ & $11.71$ & $13.71$ &$17.97$\\\hline
\end{tabular}
\caption{Logarithmic expression for energy loss rate at $\rho=
10^{15}$ $gm/cm^{3}$, and the magnetic field $H=10^{16}$,
$10^{15}$, $10^{14}$ G due to the neutrino bremsstrahlung process
in presence and absence of a magnetic field, respectively, in the
temperature range $0.8\times 10^{10} - 10^{11}$ K. The bold number
indicates that the former process dominates over the later.}
\end{table}

\pagebreak
\begin{figure}
\begin{center}
\includegraphics [scale=0.7] {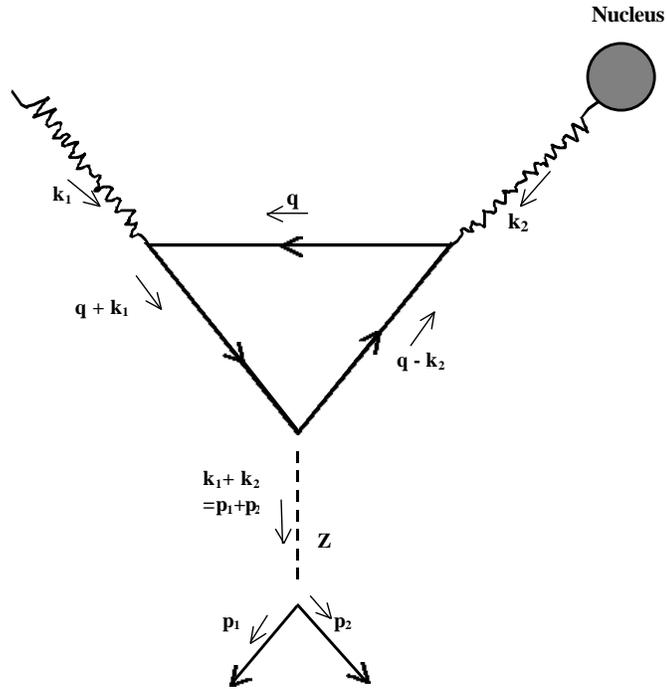}
\caption{Feynman diagram for the photo-coulomb neutrino process
with fermionic loop through the exchange of $Z$-boson.}
\end{center}
\end{figure}
\begin{figure}
\begin{center}
\includegraphics [scale=0.7] {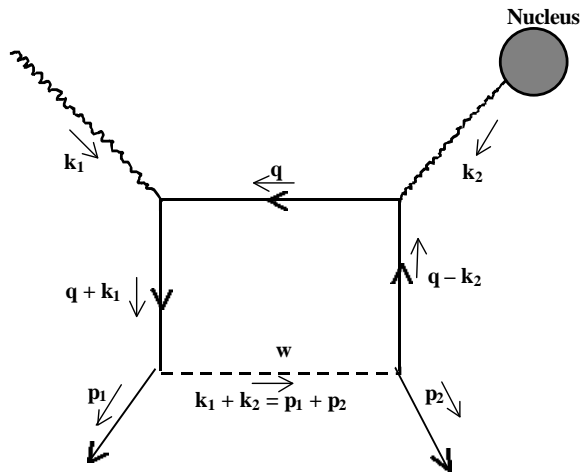}
\caption{Feynman diagram for the photo-coulomb neutrino process
having $e^{-} - W^{-} - \nu_{e}$  effect with fermionic loop.}
\end{center}
\end{figure}
\begin{figure}
\begin{center}
\includegraphics [scale=0.7] {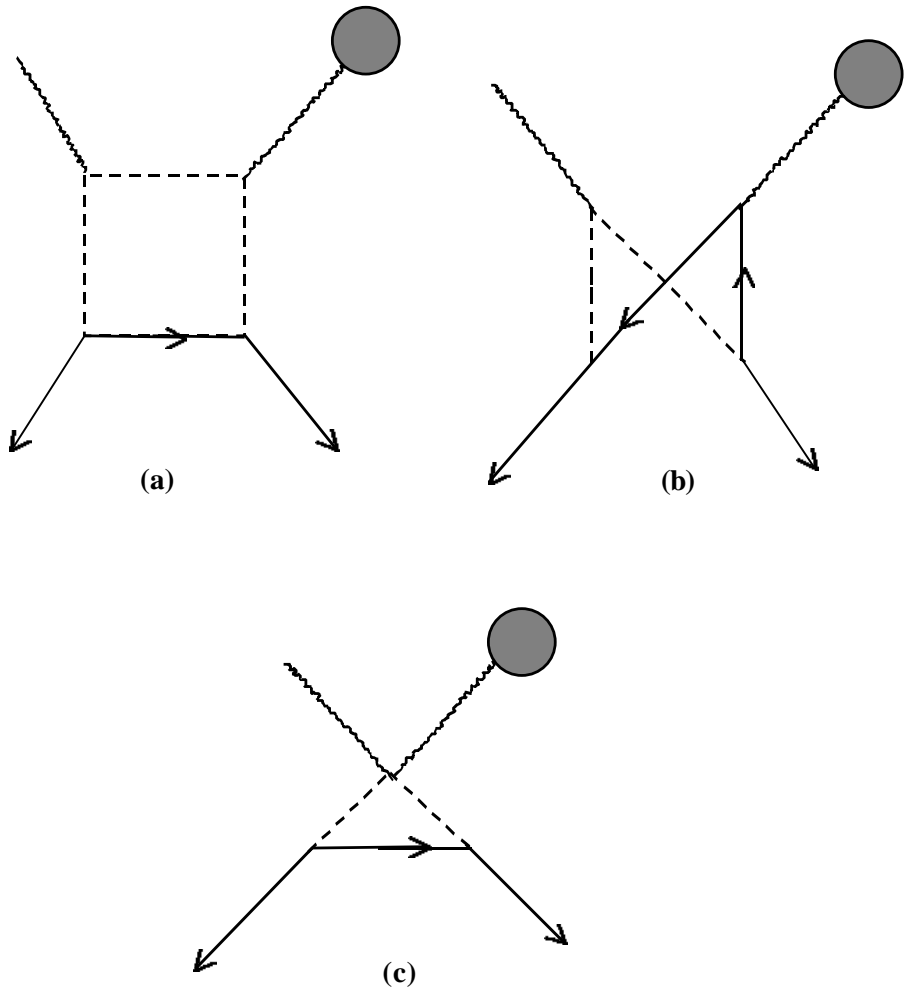}
\caption{Other diagrams for the photo-coulomb neutrino process
having $e^{-} - W^{-} - \nu_{e}$  effect in which more than one
$W$-line are present.}
\end{center}
\end{figure}
\begin{figure}
\begin{center}
\includegraphics [scale=0.7] {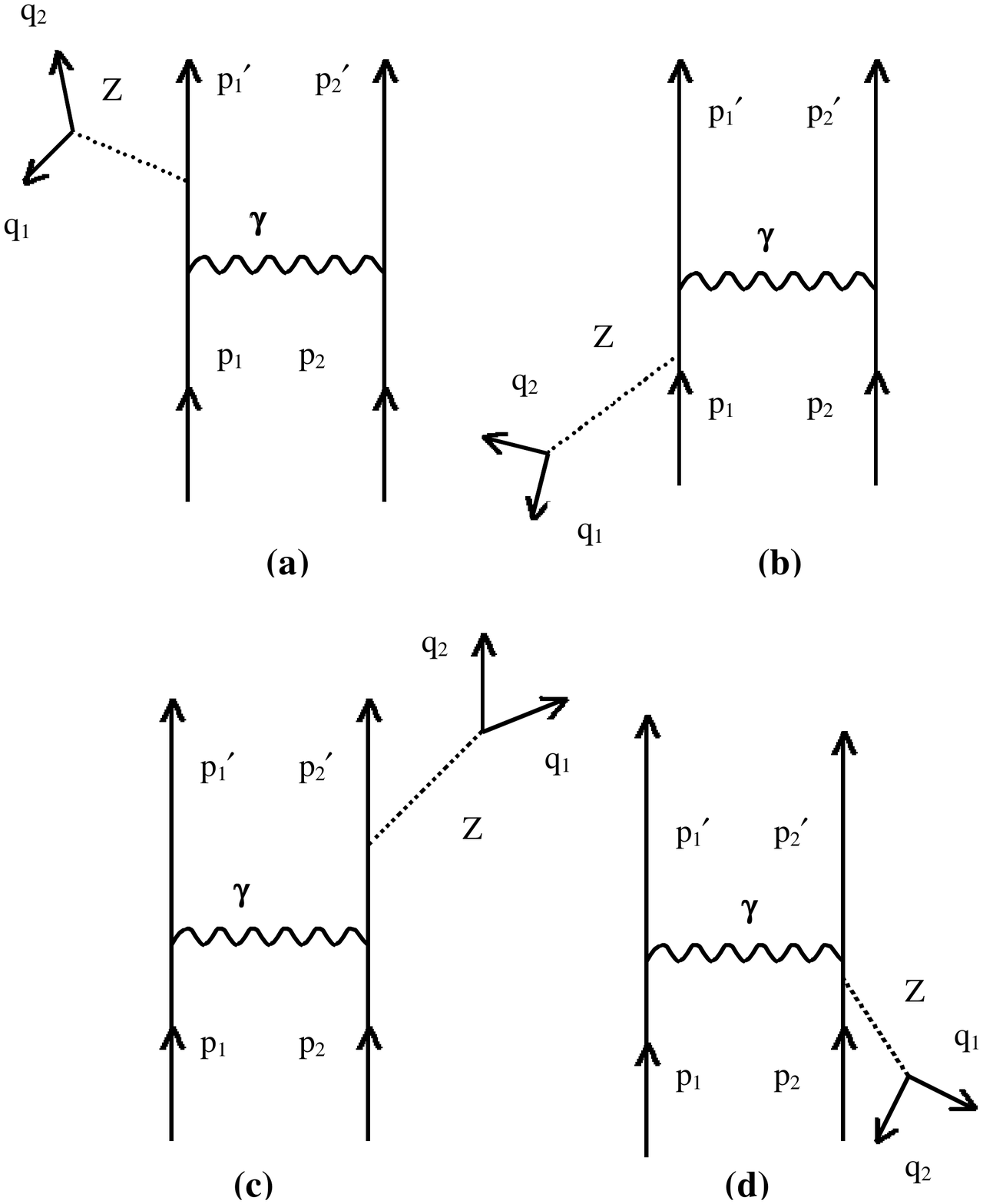}
\caption{Feynman diagrams for the direct process of
electron-neutrino bremsstrahlung.}
\end{center}
\end{figure}
\begin{figure}
\begin{center}
\includegraphics [scale=0.7] {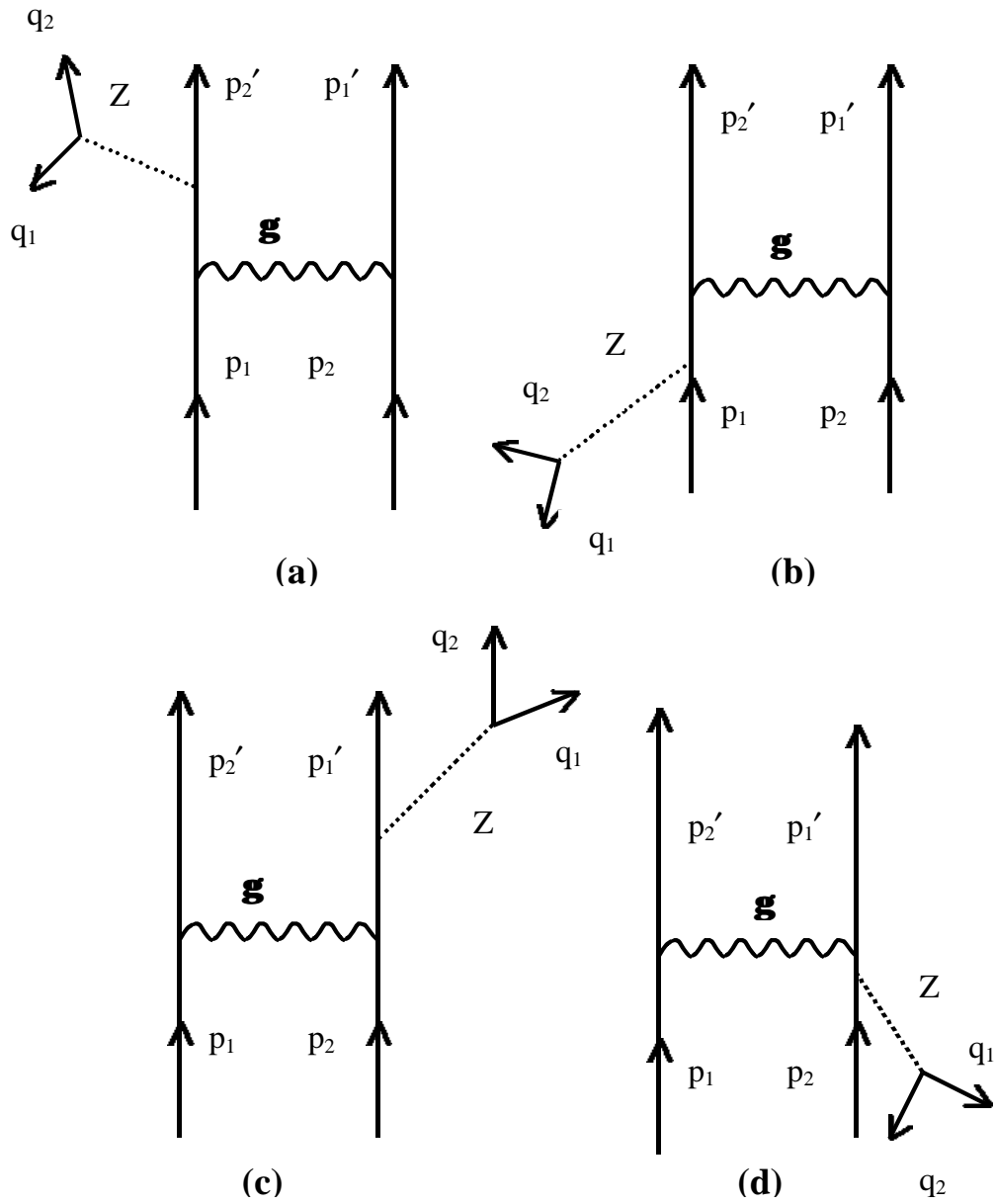}
\caption{Exchange diagrams for the electron-neutrino
bremsstrahlung.}
\end{center}
\end{figure}
\begin{figure}
\begin{center}
\includegraphics [scale=0.7] {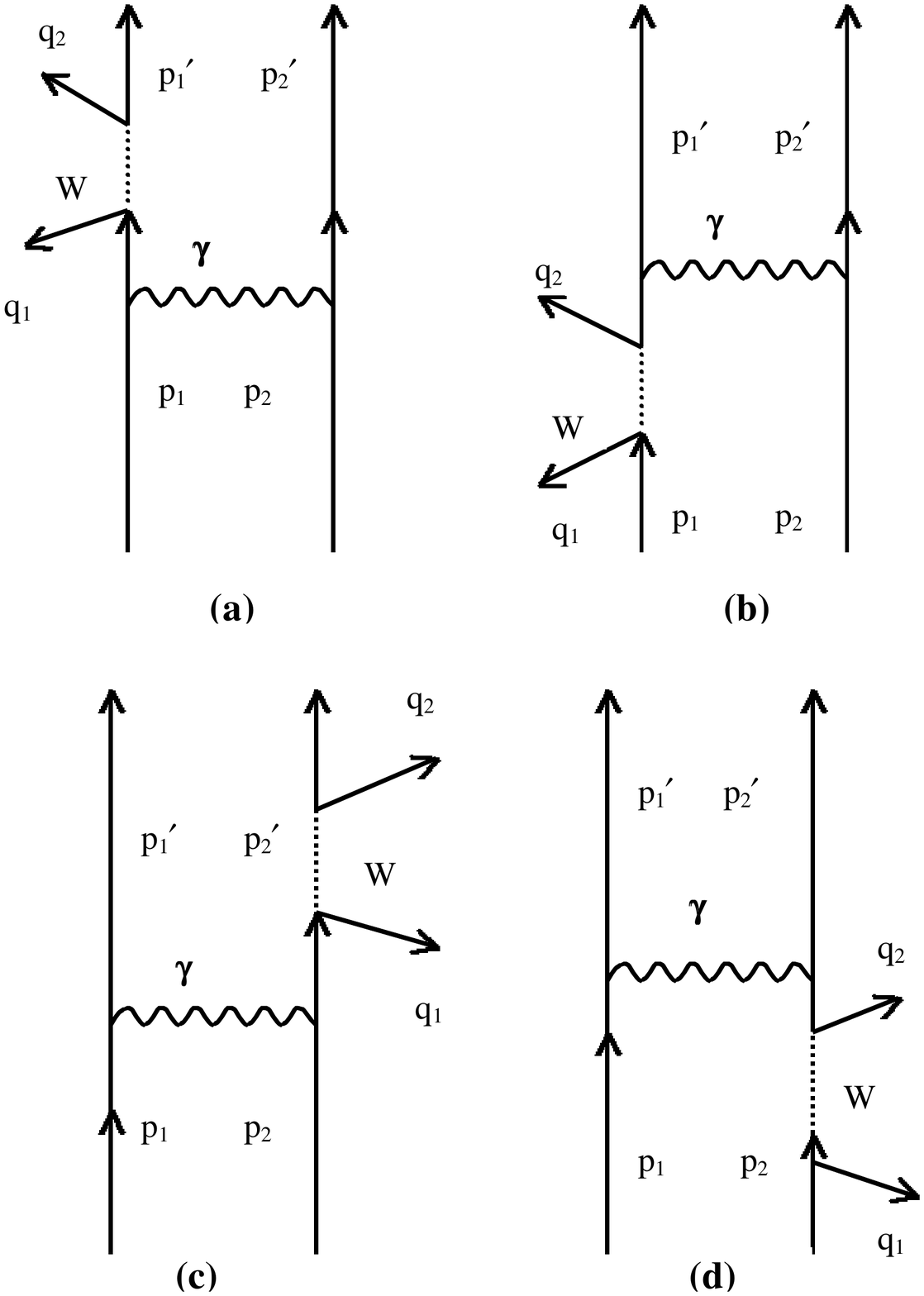}
\caption{Feynman diagrams for the direct process of
electron-neutrino bremsstrahlung having $e^{-} - W^{-} - \nu_{e}$
effect.}
\end{center}
\end{figure}
\begin{figure}
\begin{center}
\includegraphics [scale=0.7] {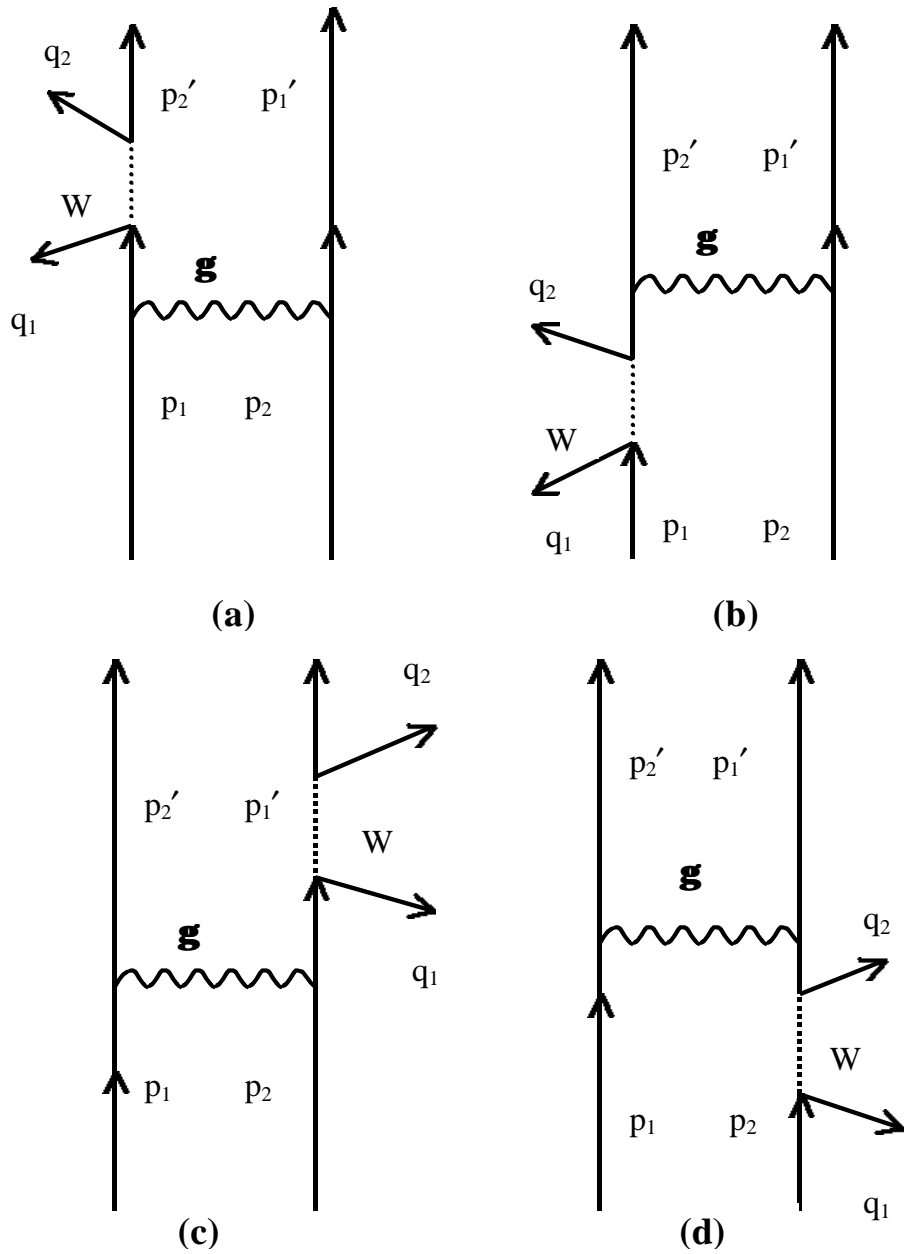}
\caption{Exchange diagrams for the electron-neutrino
bremsstrahlung having $e^{-} - W^{-} - \nu_{e}$ effect.}
\end{center}
\end{figure}
\begin{figure}
\begin{center}
\includegraphics [scale=0.7] {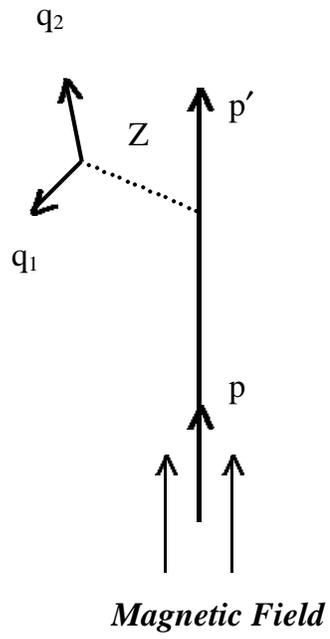}
\caption{Feynman diagram for the neutrino-synchrotron radiation}
\end{center}
\end{figure}
\begin{figure}
\begin{center}
\includegraphics [scale=0.7] {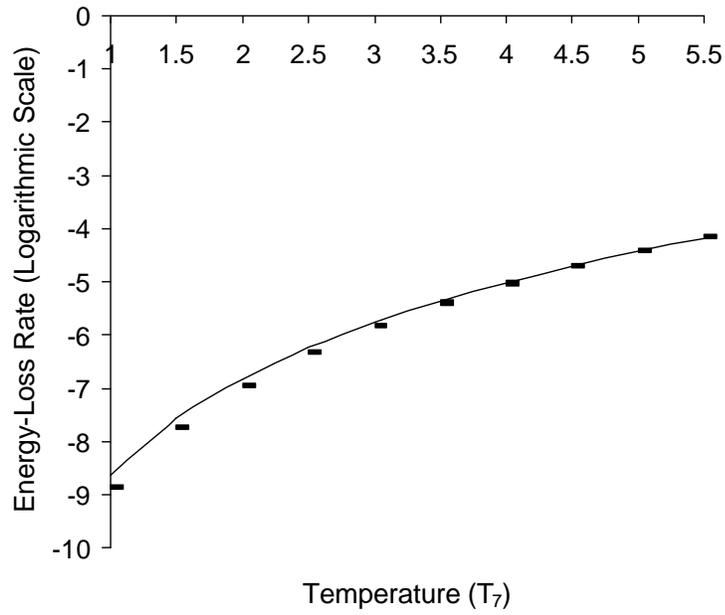}
\caption{Energy loss rate (logarithmic scale) as the function of
temperature in a white dwarf for the neutrino synchrotron
radiation, obtained in the electro-weak theory (continuous line)
and that obtained by Landstreet (dashed line) at $\rho=10^{8}$
$gm/cc$ and $H=10^{11}$ G.}
\end{center}
\end{figure}
\begin{figure}
\begin{center}
\includegraphics [scale=0.7] {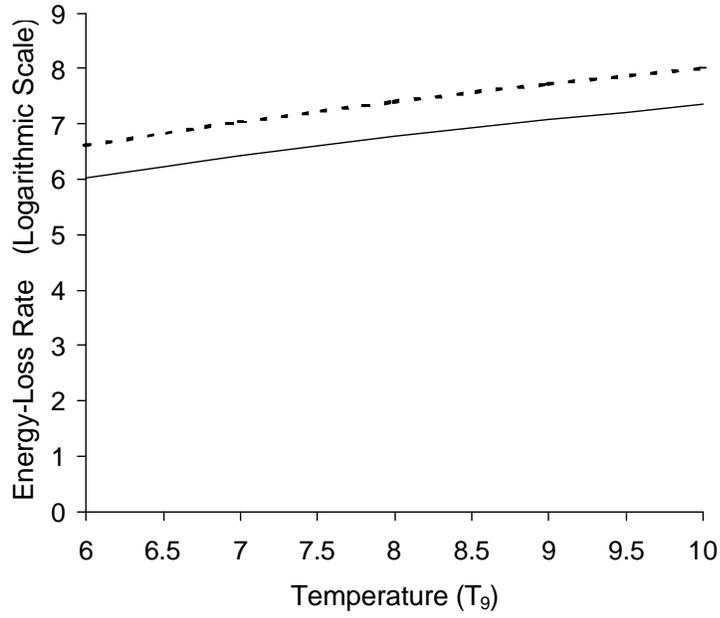}
\caption{Energy loss rate (logarithmic scale) as the function of
temperature in a neutron star for the neutrino synchrotron
radiation, obtained in the electro-weak theory (continuous line)
and that obtained by Landstreet (dashed line) $\rho=10^{15}$
$gm/cc$ and $H=H_{c}=4.414\times 10^{13}$ G.}
\end{center}
\end{figure}
\begin{figure}
\begin{center}
\includegraphics [scale=0.7] {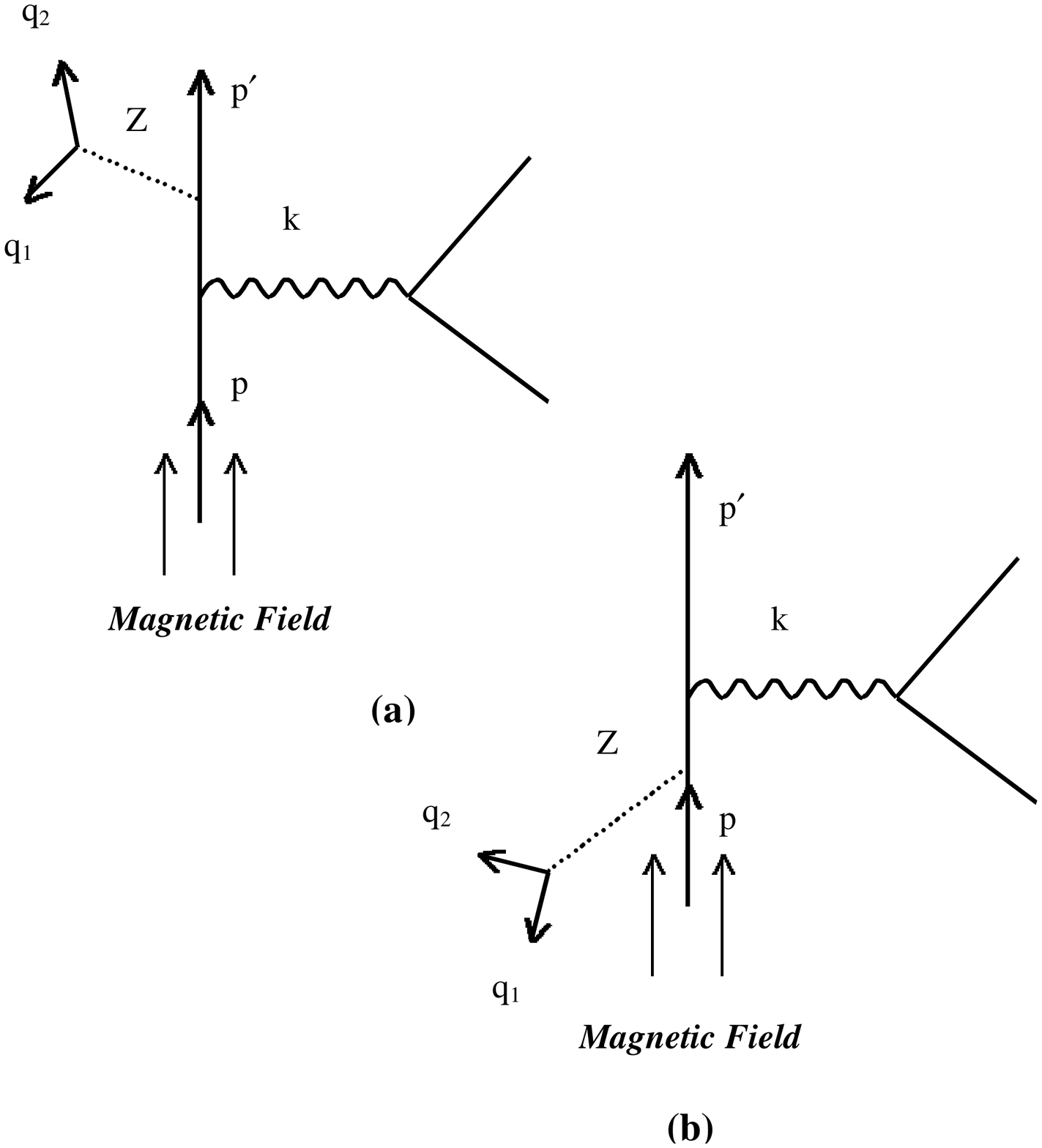}
\caption{Feynman diagrams for the neutrino bremsstrahlung process
in presence of a magnetic field.}
\end{center}
\end{figure}
\begin{figure}
\begin{center}
\includegraphics [scale=0.7] {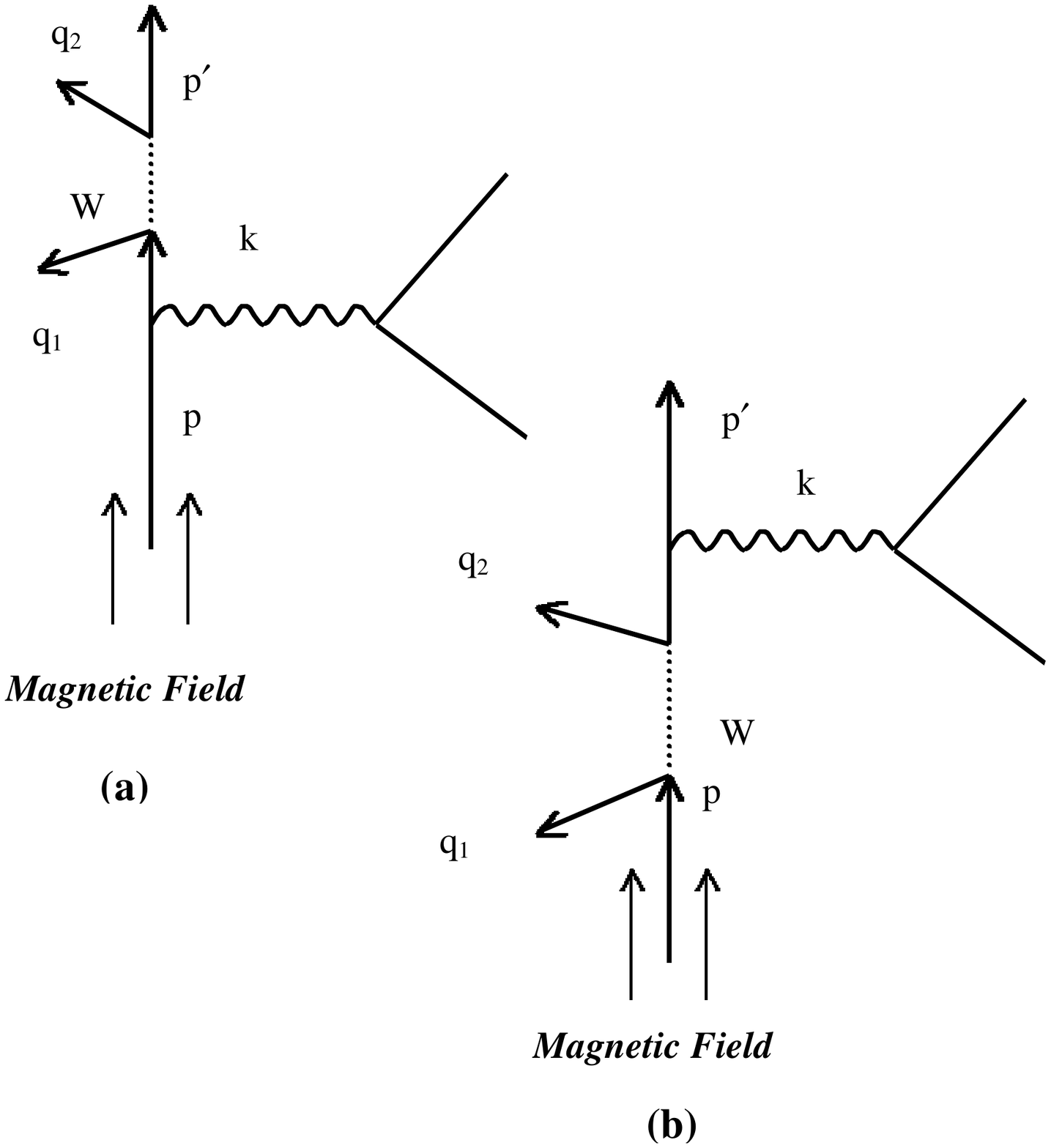}
\caption{Feynman diagrams for the neutrino bremsstrahlung process
having $e^{-} - W^{-} - \nu_{e}$ effect in presence of a magnetic
field.}
\end{center}
\end{figure}
\begin{figure}
\begin{center}
\includegraphics [scale=0.7] {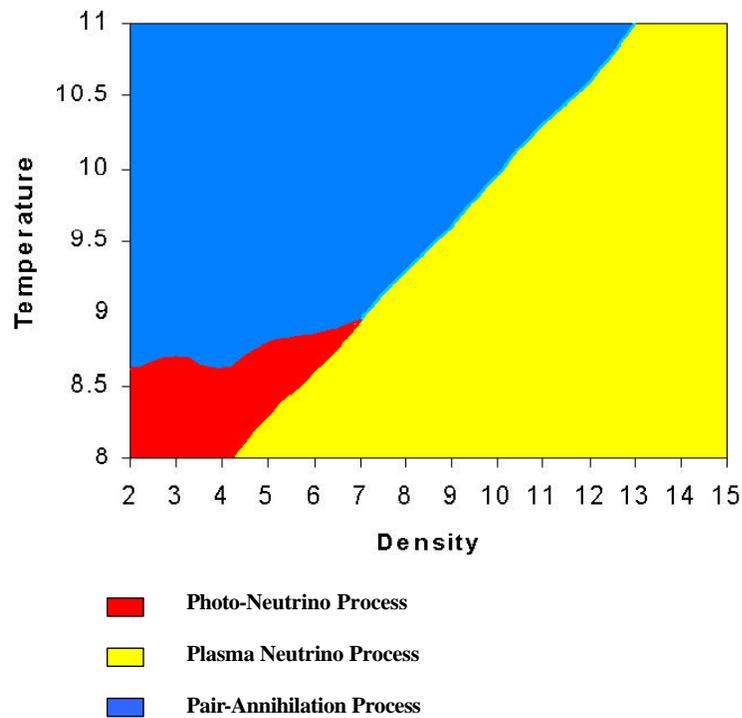}
\caption{Region of dominance of different neutrino emission
processes in absence of a magnetic field. The regions are
characterized by the density (taken along $x$-axis) and the
temperature (taken along $y$-axis), plotted in the logarithmic
scales.}
\end{center}
\end{figure}
\begin{figure}
\begin{center}
\includegraphics [scale=0.7] {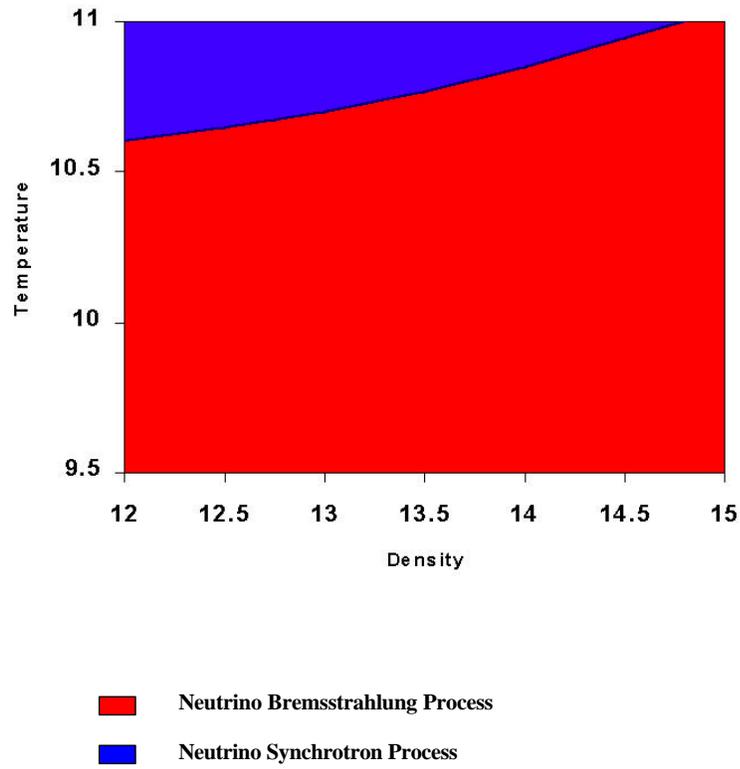}
\caption{Region of dominance of different neutrino emission
processes in presence of a strong magnetic field ($H=10^{12}$ G).}
\end{center}
\end{figure}

\end{document}